\input{epsf}
\documentstyle[prd,aps,epsf,eqsecnum,twocolumn]{revtex}

\catcode`\@=11

\def\maketitle2{\par 
\begingroup
\let\cite\@bylinecite
\def\thefootnote{\fnsymbol{footnote}}%
\twocolumn[\@maketitle2\vskip2pc]%
\thispagestyle{plain}\@thanks
\endgroup
\def\thefootnote{\arabic{footnote}}%
\setcounter{footnote}{0}%
\let\maketitle2\relax \let\@maketitle2\relax
\let\@thanks\relax \let\@authoraddress\relax \let\@title\relax
\let\@date\relax \let\thanks\relax \let\@abstract\relax 
\let\@pacs\relax}

\def\abstract#1{\gdef\@abstract{{\par 
\bgroup
\ifdim\prevdepth=-1000pt \prevdepth0pt\fi
\hsize\columnwidth
\dimen0=-\prevdepth \advance\dimen0 by17.5pt \nointerlineskip
\small\vrule width 0pt height\dimen0 \relax}{~~}#1\egroup}}

\def\pacs#1{\gdef\@pacs{{\par 
\bgroup
\hsize\columnwidth \parindent0pt
\ifdim\prevdepth=-1000pt \prevdepth0pt\fi
\dimen0=-\prevdepth \advance\dimen0 by20pt\nointerlineskip
\egroup} PACS numbers:~#1}}

\def\@maketitle2{
\@preprint
\@title
\ifdim\prevdepth=-1000pt \prevdepth0pt\fi
\@authoraddress
\@date
\begin{list}{}{\leftmargin=0.10753\textwidth \rightmargin=\leftmargin
\itemsep=1pc\partopsep=-1pc}
\item\@abstract
\item\@pacs
\end{list}
}

\catcode`\@=12

\def\ddk{[d^3{\rm k}]}
\def\kk{{\rm k}}  
\begin{document}
 
\title{Inclusive  Dilepton Production at RHIC: a Field Theory Approach Based on a Non-equilibrium Chiral Phase Transition}
\author{Fred Cooper\thanks{electronic mail:fcooper@lanl.gov}} 
\address{Theoretical Division, MS B285, Los Alamos National
Laboratory, Los Alamos, NM 87545}
\author{Submitted to the Symposium in Honor of Richard Slansky, May 20, 1998}
\author{ Dedicated to the memory of Richard Slansky, Friend and Colleague}
\date{\today}


\abstract{Recently a real time picture of quantum field theory has been
developed which allows one to look into the time evolution of a scattering
process. 
We discuss two pictures for discussing non-equilibrium processes, namely the Schrodinger Picture and the Heisenberg picture and show that a Time
Dependent Variational Method is equivalent to the leading order in Large-N
approximation in the Heisenberg picture. We then discuss the dynamics
of a non-equilibrium chiral phase transition in mean field theory in the
$O(4)$ sigma model.  We show how the pion spectrum can be enhanced at low
momentum because of non-equilibrium effects. We then show how to use
Schwinger's CTP formalism to calculate the inclusive dilepton spectrum coming
from the pion plasma. We find that a noticable enhancement occurs in this
spectrum, but that there are large numerical uncertainties due to errors connected with the finite times used to do our numerical simulations.}
  
\preprint{LAUR  98-XXXX}
\maketitle2
\section{Introduction}
\label{sec:Introduction}
In the early 70's, the study of inclusive Hadronic Interactions was at the 
forefront of theoretical research because of the opening of two major 
accelerators--FNAL in the United States and the ISR at CERN. As data
poured in many of us tried to understand from various approaches such
as Regge Poles, Fireballs, and Hydrodynamics how to determine the
single particle inclusive spectrum of particles.  In that atmosphere,in 1974,
 Peter
Carruthers founded the Particle Physics group at Los Alamos which initially
included Dick Slansky, Geoff West, David Campbell, David Sharp, Mitch 
Feigenbaum and myself, all of whom were working on this problem when we
first arrived.  At that time QCD was not yet a well developed subject, computers were relatively archaic and
the idea of a first principles approach to understanding what happens
after two particles collide was just a dream.  Recently this dream has
gotten closer to reality, and as a tribute to Dick and our early efforts
I would like to review progress we have made in this direction.  
I will begin by  first reviewing two pictures for discussing non-equilibrium
processes, namely the Schrodinger and Heisenberg Pictures.  In the 
Schrodinger Picture one can reduce the number of degrees of freedom by using
a time-dependent variational method with a trial density matrix which is Gaussian.
In the Heisenberg picture one can use the large N expansion of the Path Integral
in the Closed Time Path Formalism to give us a controllable expansion
about the mean-field approximation. We will then review our model for
a non-equilibrium phase transition, namely the O(4) linear sigma model, 
and set up the time evolution equations for the lowest order in Large $N$
approximation.  In this model we will obtain natural cooling (quenching)
of the plasma from the unbroken phase to the broken symmetry phase as a result
of expansion into the vacuum with boost invariant kinematics imposed.  
We find that there is a growth of unstable modes when the order parameter
(effective pion mass) goes negative during the evolution.  This causes a distortion of the single particle distribution of pions.  We also reconstruct typical
classical configurations by sampling the quantum density matrix and see domains. We then
obtain the dilepton spectrum from this time evolving plasma using Schwinger's
closed time path formalism.

\section{Strategies for Studying Time Evolution Problems in $\lambda\varphi^{4}$ Field Theory}
\subsection{Schrodinger Picture}
In quantum mechanics, time evolution problems are usually discussed in the 
Schrodinger Picture. The initial state in the x representation is the wave function
\begin{equation}
\Psi(x,t) = < x \vert \Psi>
\end{equation}
which evolves in time according to the Schrodinger equation:
\begin{equation}
i {\partial \Psi(x,t) \over \partial t}  = H \Psi(x,t)
\end{equation}
This equation can be obtained from Dirac's Variational Principle \cite{Dirac}. Define
the action
\begin{equation}
\Gamma = \int dt < \Psi | i \partial/\partial t -H| \Psi >
\end{equation}
Minimimizing the Action then yield the Schrodinger equation: 
\begin{equation}
 \delta\Gamma = 0 \rightarrow  \{ i {\partial \over \partial t} -H \} |\Psi > = 0 
\end{equation}
Here one thinks of $H$ as an operator which in the $x$ representation one has
\begin{equation}
p= -i {d \over dx}
\end{equation}
Gaussian initial conditions for the wave function would be:
\[ \Psi(x,0) = \exp [-\alpha(x-x_{0})^{2}]. \]

In field theory, the wave function is replaced by a wave functional.
 For example
a Gaussian Wave functional is given by:
\begin{eqnarray}
&&<\varphi | \Psi> = \psi[\varphi, t] =  \nonumber \\  
&&\exp[[-\int_{x,y}[\varphi(x)-\hat{\varphi}(x)]  [{G^{-1}(x,y) \over
4}-i\Sigma(x,y)][\varphi(y)-
\hat{\varphi}(y)]  \nonumber \\
&&+ i \hat{\pi}(x) [\varphi(x)-\hat{\varphi}(x)]] \label{varwave}  
\end{eqnarray}

The time evolution is now given by the  functional Schrodinger equation :

\begin{eqnarray}
i {\partial\psi \over \partial t} &=& H\psi \nonumber \\ 
H &=& \int d^{3}x [- {1 \over 2}
\delta^{2}/\delta \varphi^{2}+ {1 \over 2} (\nabla\varphi)^{2} + V(\varphi)]
\end{eqnarray}
 
Dirac's variational principle 
\begin{equation}
\Gamma = \int dt < \Psi | i \partial/\partial t -H| \Psi >
\end{equation} 
 \[ \delta\Gamma = 0 \rightarrow  \{ i {\partial \over \partial t} -H \} |\Psi > = 0 \]
is the starting point for thinning the degrees of freedom. Instead of 
putting the functional Schrodinger equation on the computer, one assumes that
the wave functional stays Gaussian, and uses the variational principle to
determine the time evolution equations for the one and two point functions
involved in the description of the Gaussian. This approximation is 
equivalent to mean field theory and at large-N becomes equivalent to the leading order
large-$N$ approximation. That is we assume a 
Gaussian 
trial wave functional: 
\begin{eqnarray}
&&<\varphi|\Psi_{v}> =\psi_{v}[\varphi,t]
\end{eqnarray}
with $\psi_v$ given by Eq.(\ref{varwave})
The variational parameters have the following meaning:
\begin{eqnarray}
\hat{\varphi}(x,t) &=& < \Psi_{v}|\varphi| \Psi_{v} >; \hat{\pi}(x,t) =
<\Psi_{v}|-i \delta/\delta\varphi| \Psi_{v} >
\nonumber \\
 G(x,y,t) &&= < \Psi_{v}|
\varphi(x)\varphi(y)| \Psi_{v} > - \hat{\varphi}(x,t) \hat{\varphi}(y,t) 
\end{eqnarray}
The Effective Action for the variational parameters is
\begin{eqnarray}
&&\Gamma(\hat{\varphi}, \hat{\pi},G,\Sigma) =\int dt < \Psi_{v}| i
\partial/ \partial t -H| \Psi_{v}>  \nonumber \\
&&= \int dt dx[\pi(x,t)\dot{\varphi}(x,t) +\int dt dx dy \Sigma(x,y)\dot{G}(x,y,t)  \nonumber \\
&&-\int dt < H > 
\end{eqnarray}
where
\begin{eqnarray}
< H >&& = \int dx \{{\pi}^{2}/2 + 2 \Sigma G\Sigma + G^{-1}/8 + 1/2 
(\nabla \varphi)^{2} \nonumber \\
&& -1/2 \nabla^{2}G   + 1/2 V''[\varphi] G + 1/8 V''''[\varphi]
G^{2}\}.
\end{eqnarray}

Equations of motion that result from varying the action are:
\begin{eqnarray}
\dot{\pi}(x,t) &&= \nabla^{2}\varphi - \partial <V> /\partial\varphi;  \nonumber \\
\dot{\varphi}(x,t) &&= \pi  \nonumber \\
\dot{G}(x,y,t) &&= 2\int dz[\Sigma (x,z)G(z,y)+G(x,z)\Sigma (z,y)]  \nonumber \\
\dot{\Sigma}(x,y,t)&&= -2 \int dz[\Sigma (x,z)\Sigma (z,y) + G^{-2}/8 \nonumber \\
&& +  [{1 \over 2}\nabla^{2}_{x} - \partial <V> / \partial G] \delta^{3}(x-y) 
\end{eqnarray}
In $\phi^4$ theory if there is translational invariance, we can simplify these
equations by Fourier transforming them in three dimensional spase to obtain:

\begin{eqnarray}
 &&2 \ddot{G} (k,t)G(k,t) - \dot{G}^{2}(k,t) +4 \Gamma (k,t)G^{2}(k,t) - 1 = 0  \nonumber \\
&&\Gamma (k,t) = k^{2}+m^{2}(t); m^{2}(t) = -\mu^{2} + {1 \over 2}\lambda \int [dk] G(k,t) \nonumber \\
\end{eqnarray}
One can also linearize these equations by recognizing that the mean field
approximation for the homogeneous problem is equivalent to a field theory
with a time dependent mass which is self consistently determined. That is
if we assume a quantum field obeying:
\begin{equation}
({\Box + m^2(t)} )\phi(x,t) = 0
\end{equation}\
then one can satisfiy the equation for $G(x,y,t)$ by choosing:
\begin{equation}
G(x,y;t) \equiv < \phi(x,t) \phi(y,t) >
\end{equation}

\subsection{Heisenberg Picture--Path Integral Approach-- Closed time-path approach of J. Schwinger}

In the Heisenberg Picture, the operators are time dependent, and the expectation value of the fields in an initial state are the infinite number of 
c-number variables. The infinite heirarchy of coupled Green's functions
need to be truncated by an approximation scheme. The large $N$ approximation
orders the connected Green's functions in powers of $1/N$, with the 
connected four point function going as $1/N$, 6 point function $1/N^2$ etc.
The formalism for preserving causality in initial value problems was invented
by Schwinger \cite{ref:SchKel} and was later cast in the form of path integrals
\cite{ref:Zhou}. 
The
Generating Functional for initial value problem Green's functions is
\begin{eqnarray}
 Z [J^{+}, J^{-}, \rho]&& =\int d \Psi_{out}< i| T^{\ast}(\exp \{ - \int
iJ_{-}\varphi_{-}\} )|\Psi_{out} > \times \nonumber \\
&& < \Psi_{out} | T(\exp
\int iJ_{+}\varphi_{+})| i >  \nonumber \\
&&= Tr \{ \rho   T^{\ast}(\exp \{ - \int
iJ_{-}\varphi_{-}\} ) T(\exp
\int iJ_{+}\varphi_{+}) \} \nonumber \\
\end{eqnarray}

This can be written as the product of an ordinary Path integral 
times a complex conjugate one or as
a matrix Path integral.
\begin{eqnarray}  
&& Z [J^{+}, J^{-}]= \int \prod_{+,-} d \varphi_i  Tr \rho \{
\exp i[(S[\varphi_{+}] + \int J_{+}\varphi_{+}) \nonumber \\
&& -
(S^{\ast}[\varphi_{-}] + \int J_{-}\varphi_{-}) ]  \}\nonumber \\
 &= & \int \prod_{\alpha}  d\varphi_{\alpha} \exp i
(S[\varphi_{\alpha}] + J_{\alpha}\varphi^{\alpha} ) 
< \varphi_{1},i|\rho |\varphi_{2} , i > \nonumber \\
&& \equiv e^{i W[J_{\alpha}]}
\end{eqnarray}
where  
$< \varphi_{1},i|\rho |\varphi_{2} , i >$ is the density matrix defining the initial state.
We use the matrix notation:
\begin{equation} 
\varphi^a = \left(\begin{array}{c} \varphi_+ \\ \varphi_- \end{array}\right)\ ;
\qquad a=1,2\
\end{equation}
with a corresponding two component source vector,
\begin{equation}
J^a = \left(\begin{array}{c} J_+ \\
 J_- \end{array}\right)\ ;
\qquad a=1,2\ .
\end{equation}
On this matrix space there is an indefinite metric
\begin{equation} c_{ab} = diag \ (+1,-1) = c^{ab} \label{metr}
\end{equation}
 so that, for example  
\begin{equation} 
J^ac_{ab}\varphi^b = J_+ \varphi_+ - J_-\varphi_-\ .  
\end{equation}

From the Path Integral we get the following Matrix Green's Function:

\begin{equation}
 G^{ab}(t,t') = {\delta^{2} W \over \delta
J_{a}(t) \delta J_{b}(t')} \bigg\vert_{J =0}\ . 
\end{equation} 
 \begin{eqnarray} G^{21}(t,t') &\equiv& G_> (t,t') = \ i{\rm
Tr}\{\rho\ \varphi(t) \overline{\varphi}(t') \}_{con}\ ,\nonumber \\
 G^{12}(t,t') &\equiv& G_<
(t,t') = \pm i{\rm Tr}\{\rho\ \overline{\varphi}(t') \varphi(t) \}_{con} \nonumber \\
G^{11}(t,t') &=& 
i{\rm Tr}\left\{\rho {\cal{T}} [\varphi (t) \overline{\varphi}(t') ] \right\}_{con}\nonumber \\
&& = \Theta (t,t')
G_> (t,t') + \Theta (t',t) G_< (t,t')\nonumber \\
 G^{22}(t,t') &=&  i{\rm Tr}\left\{\rho\
{\cal{T}}^* [\varphi(t) \overline{\varphi}(t')] \right\}_{con}\nonumber \\
&&= \Theta (t',t) G_> (t,t') + \Theta
(t,t') G_< (t,t') \nonumber \\
\label{matr} 
\end{eqnarray}
We notice that $ G_F=G^{11}(t,t')$ and $G_{F^*}= G^{22}(t,t')$. \\

We also will need the relationships:
\begin{eqnarray}
 G_{ret}(t,t')&& = i \Theta (t-t') [ \Phi(t), {\bar \Phi}(t')]_{\pm}\nonumber\\
&& =  \Theta (t-t') [G^>(t,t') -G^<(t,t')].
\end{eqnarray}
\begin{eqnarray}
 G_{adv}(t,t')&& = -i \Theta (t'-t) [ \Phi(t), {\bar \Phi}(t')]_{\pm} \nonumber \\
&& =
  \Theta (t'-t) [G^<(t,t') -G^>(t,t')].
\end{eqnarray}
as well as the relations between the Green's functions:
\begin{eqnarray}
G_{ret}(t,t')&&= G_F(t,t')-G^<(t,t') = -G_{F^*}(t,t') + G^>(t,t') \nonumber \\
G_{adv}(t,t')&&= G_F(t,t')-G^>(t,t') = -G_{F^*}(t,t') + G^<(t,t') \nonumber \\
\end{eqnarray}

\subsubsection{Large-N Approximation} 
The method for reducing the number of degrees of Freedom in the Heisenberg picture is the
 large-$N$ approximation \cite{largeN}
If we have an $N$-component scalar field with Lagrangian:
\begin{eqnarray}
\tilde L_{cl}[\Phi] &=& \frac{1}{2} (\partial_{\mu}\Phi_i)(\partial^{\mu}
\Phi_i) -{\lambda\over 8N}\left(\Phi_i\Phi_i - {2N\mu^2\over\lambda}\right)^2
\nonumber\\   
\label{lag1} 
\end{eqnarray}
we can rewrite this as:
\begin{eqnarray}
\tilde L_{cl}[\Phi,\chi] &&=
-\frac{1}{2} \Phi_i (\Box+ \chi) 
\Phi_i +
{N\over \lambda} \chi\left({\chi \over 2}  +  \mu^2
\right)\nonumber\\ 
\label{lag}
\end{eqnarray}
where $i = 1, \ldots , N$ and
\begin{equation} 
\chi  = -\mu^2 + {\lambda\over 2N} \Phi_{i} \Phi_{i}~,  
\label{chi}
\end{equation}

If  $\mu^2 > 0$, spontaneous symmetry breaking at the classical level. 
At this minimum the $O(N)$ symmetry is spontaneously broken, $\chi =0$
and there are $N-1$ massless modes. Small oscillations in the
remaining $i=N$ (radial)  direction describe a massive mode with bare mass equal
to $\sqrt 2 \mu = \sqrt\lambda v_0$.
The Generating functional for all Graphs is given by \cite{cgs}:
\begin{equation}
Z[j,K] = \int d\phi d\chi \exp \{i S[ \phi,\chi]+  i \int [j \phi + K \chi] \}
\end{equation}

Perform the Gaussian integral over the field $\phi$
\begin{equation}
Z[j,K]= \int d \chi \exp  \{i N S_{eff} [\chi, j,K] \}
\end{equation}
where
\[
 S_{eff} =\int dx \{ {1 \over 2} j G^{-1}[\chi] j +  K \chi +{1\over \lambda}
\chi\left({\chi \over 2}  +  \mu^2 \right) + {i \over 2} \rm{Tr}  \ln G^{-1}
[\chi] \} \]
\begin{equation}
G^{-1}[\chi](x,y)  \equiv \{ \Box  + \chi\ \} \delta (x-y) ,
\label{Ginv}
\end{equation}
Because of the $N$ in the exponent one is allowed to perform
the integral over $\chi$ by stationary phase. This leads
to an expansion of $Z$ in powers of $1/N$ the lowest term
(stationary phase point) is related to the previous Gaussian  (Hartree)
approximation.  The effective action of the leading order is:

\begin{equation}
{\cal S}_{eff} [\phi, \chi] =  S_{cl}[\phi, \chi] + 
{\frac{i\hbar}{2}}  \ {\rm Tr} \ln G^{-1}[\chi]\ . 
\label{Seff}
\end{equation}
Varying the action leads to the mean field equations:
\begin{equation}
\{\Box + \chi \} \phi =0
\end{equation}
\[  \chi =  - \mu^2 + {\lambda \over 2 N} ( \phi^2 + {1 \over i} G(x,x;\chi) ) \]

We notice that this is the same equation found in the Gaussian approximation
with $m^2(t)$ being identified with $\chi$. 

\section{ Dynamical evolution of a non-equilibrium Chiral Phase Transition}

One important question for RHIC Experiments is-- can one produce disoriented
chiral condensates (DCC's)  in a relativistic heavy ion collision?
Recently, Bjorken, 
Rajagopal and Wilczek and others  proposed that a nonequilibrium chiral phase
transition such as a quench might lead to 
regions of DCC's \cite{bib:dcc1}.
The model Rajagopal and Wilczek  considered was  the $O(4)$ linear sigma model in a tree-level
approximation, where a quench was assumed.  Two deficiencies of that
model were its classical nature (it could not describe $\pi$-$\pi$ scattering
), and the quench was put in by hand.  Our approach \cite{bib:dcc} instead was to look
at the quantum theory in an approximation that captures the phase structure
as well as the low energy pion dynamics. We also used the natural expansion
of an expanding plasma to cool the plasma and built into our approximation
boost invariant kinematics which result from a hydrodynamic picture where
the original plasma is highly Lorentz contracted.In the linear sigma model
treated in leading order in the $1/N$ expansion the theory has a chiral 
phase transition at around 160 Mev and we choose the parameters of this theory to  give a
reasonable fit to the correct low energy scattering data. We obtain natural quenching for
certain initial conditions as a result of the expansion process. 

\subsection{review of the linear $ \sigma$ model}
The Lagrangian for the O(4) $\sigma$ model is:

\begin{equation}
L= {1\over 2} \partial\Phi \cdot \partial\Phi - {1\over 4} 
\lambda (\Phi \cdot \Phi - v^2)^2 + H\sigma.
\end{equation}
The mesons  form an $O(4)$ vector
 \[\Phi = (\sigma, \pi_i)\]
As we discussed earlier in our discussion of the large-$N$ approximation we
introduce:
\[\chi = \lambda (\Phi \cdot \Phi-v^2)\]
and use the equivalent Lagrangian:
\begin{equation}
L_2 = -{ 1 \over 2} \phi_i (\Box + \chi) \phi_i + {\chi^2 \over 4
\lambda} + 
{1 \over 2} \chi v^2 + H \sigma
\end{equation}

The leading order in $1/N$ effective action which we obtain by integrating
out the $\phi$ field and keeping the stationary phase contribution to the
$\chi$ integration is

\begin{equation}
\Gamma[\Phi,\chi] = \int d^4x[ L_2(\Phi,\chi,H) + { i \over 2}N {\rm
tr~ln}
G_0^{-1}]
\end{equation}
$$
G_0^{-1}(x,y) = i[\Box + \chi(x)] ~\delta^4(x-y)
$$
This results in the equations of motion:
\begin{equation}
[\Box + \chi(x)] \pi_i = 0 ~~~~ [\Box + \chi(x)]\sigma = H,
\end{equation}
and the constraint or gap equation:
\begin{equation}
\chi= - \lambda v^2 + \lambda (\sigma^2 + \pi \cdot \pi) + \lambda
N  G_0 (x,x). 
\end{equation}
We will introduce fluid proper time and rapidity variables to implement the
kinematic constraint of boost invariance.
$$\tau\equiv(t^2-z^2)^{1/2}, \qquad \eta\equiv{1\over 2} 
\log({{t-z}\over{t+z}}).
$$
To implement boost invariance we assume that 
mean  (expectation) values of the fields     $\Phi$  and $\chi$ are
functions 
of $\tau$ only. 
\begin{eqnarray}
&&\tau^{-1}\partial_\tau\ \tau\partial_\tau\ \Phi_i(\tau)
 +\ \chi(\tau)\ \Phi_i(\tau) =
 H \delta_{i1} \nonumber \\
&&\chi(\tau) =\lambda\bigl(-v^2+\Phi_i^2(\tau)+
N G_0(x,x;\tau,\tau)\bigr),   
\end{eqnarray}
To calculate the 
 Green's function $G_0(x,y;\tau,\tau^{\prime})$ we first determine the
 auxiliary quantum field $\phi(x,\tau)$ 
\begin{equation}
\Bigl(\tau^{-1}\partial_\tau\ \tau\partial_\tau\ -
\tau^{-2}\partial^2_\eta  
-\partial^2_\perp + \chi(x)\Bigr)
\phi(x,\tau)=0.
\end{equation}
$$
    G_0 (x,y;\tau, \tau^{\prime}) \equiv <T\{ \phi(x,\tau) ~\phi(y,\tau^{\prime})\}>.
$$

We expand the quantum field in an orthonormal basis: 
$$
\phi(\eta,x_\perp,\tau)\equiv{1\over{\tau^{1/2}}} \int \ddk\bigl(\exp(ik
x)
f_\kk(\tau)\ a_\kk\ + h.c.\bigr)
$$
where $k x\equiv k_\eta \eta+\vec k_\perp \vec x_\perp$, 
$\ddk\equiv dk_\eta d^2k_\perp/(2\pi)^3$. 
The mode functions and $\chi$  obey:
\begin{equation}
\ddot f_\kk + 
\bigl({k_\eta^2\over{\tau^2}}+\vec k_\perp^2 + \chi(\tau) +
{1\over{4\tau^2}}\bigr) 
f_\kk=0.\label{eq:mode}
\end{equation}
\begin{equation}
\chi(\tau) = \lambda\Bigl(-v^2+\Phi_i^2(\tau)+{1\over \tau}N \int \ddk
|f_\kk(\tau)|^2\  (1+2\ n_\kk) \Bigr) \label{eq:chi}.
\end{equation}

 when $\chi$ goes negative, the low momentum modes with
$${k_\eta^2 +1/4 \over{\tau^2}}+\vec k_\perp^2  < | \chi |$$
grow exponentially.
These growing  modes then feed back into the $\chi$
equation and get damped. 
Low momementum growing
modes  lead to the possiblity of DCC's as well
as a modification of the low momentum distribution of particles. 
To fix the parameters of this mode we use the PCAC relation:
\begin{equation}
\partial_{\mu} A_{\mu}^i (x) \equiv f_{\pi} m_{\pi}^2 \pi^i(x) = H  \pi^i(x).
\end{equation}
and the definition of the broken symmetry vacuum.
\[ \chi_0 \sigma_0= m_{\pi}^2 \sigma_0=H\]
 \[ \sigma_0=f_{\pi} = {\rm {92.5 MeV} } \]  
$$
m_{\pi}^2 = - \lambda v^2 + \lambda f_{\pi}^2 +
 \lambda N \int_0^{\Lambda} \ddk {1\over 2\sqrt{k^2+m_{\pi}^2}}.
$$
The mass renormalized gap equation is
\begin{eqnarray}
&&\chi(\tau)-m_{\pi}^2 = -\lambda f_{\pi}^2+ \lambda \Phi_i^2(\tau) \nonumber \\
&&+{\lambda \over \tau}N
\int \ddk
\{|f_\kk(\tau)|^2\  (1+2\ n_\kk) - {1\over 2\sqrt{k^2+m^2}}. \}\nonumber \\
\end{eqnarray}
$\lambda$ is chosen to fit low energy scattering data\\ 
We choose our initial data
(at $\tau_0 = 1$)so that the system is in local thermal equilibrium in a comoving frame 
\begin{equation}
n_k = {1 \over e^{\beta_0  E^0 _k} -1}
\end{equation}
where $ \beta_0 = 1/T_0$ and $E^0_k=
\sqrt{{k_\eta^2\over{\tau_0^2}}+\vec k_\perp^2 + \chi(\tau_0)}$.

The initial value of
$\chi$ is determined by the equilibrium gap equation for an initial
temperature of $ 200 MeV$ and is $.7 fm^{-2}$ and the initial value of $\sigma$ is just ${H \over \chi_0 }$. The phase transition in this model occurs at a critical
temperature
of $160 MeV$.\\
To get into the unstable domain, we then introduce fluctuations in the time derivative of the classical field. \\
For $\tau_0 = 1 fm$ there is a narrow range of initial
values that
lead to the growth of instabilities
$.25 <  \vert \dot{\sigma} \vert   < 1.3$.

The results of numerical simulations described in \cite{bib:dcc} for the 
order parameter $\chi$ are shown in figure 1.

\epsfxsize=7cm
\epsfysize=6cm
\centerline{\epsfbox{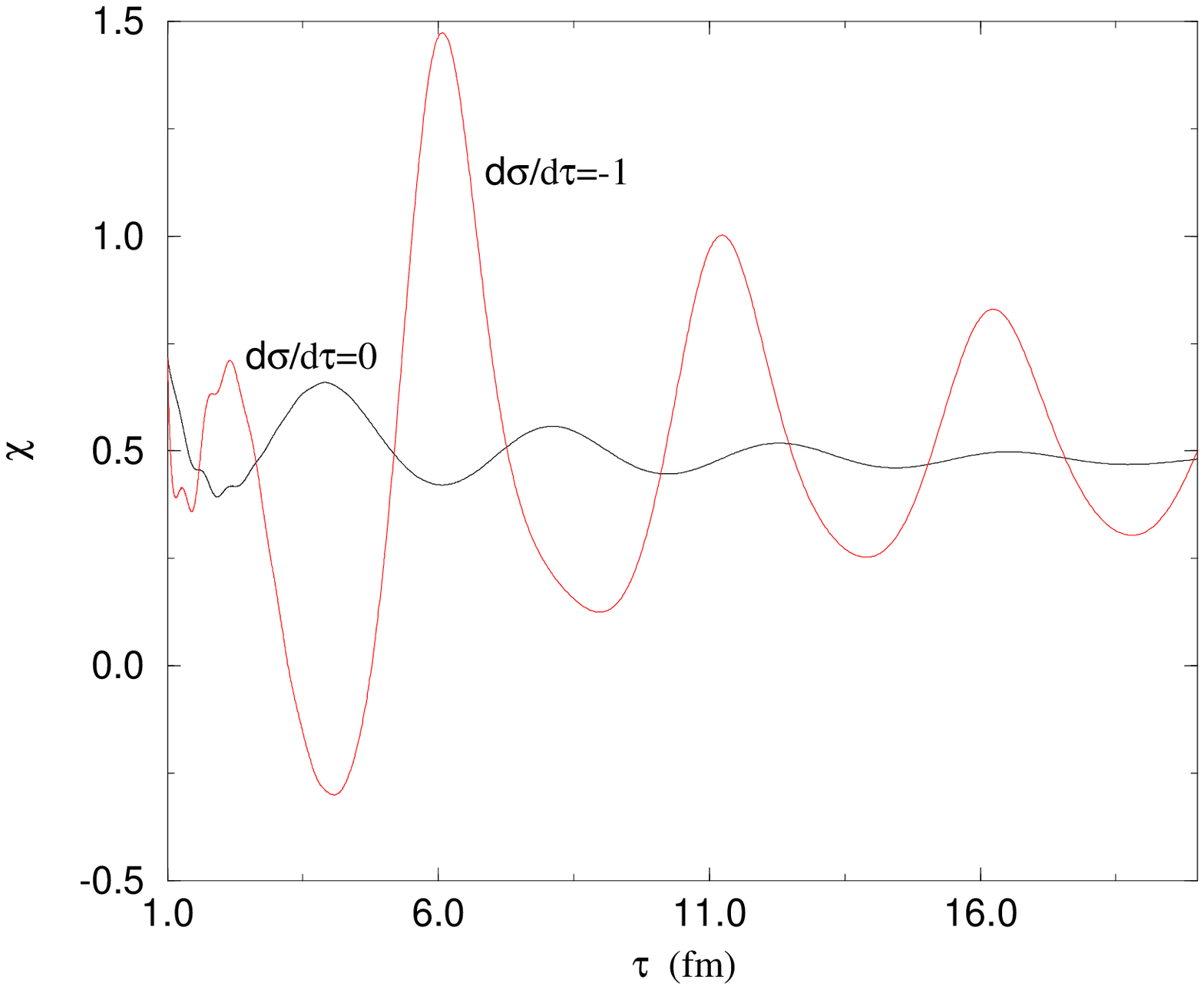}}
\vspace{.35cm}

{FIG. 1.{\small{Proper time evolution of the $\chi$ field for two
different initial values of $\dot{\sigma}$.}}}\\

Fig.1  displays the results of the numerical simulation for the
evolution of $\chi$ (\ref{eq:mode})--(\ref{eq:chi}).
We display the auxiliary field $\chi$ in units of $fm^{-2}$ 
, the classical fields $\Phi$ in units of $fm^{-1}$ 
and the proper time
in units of $fm$   ($ 1fm^{-1} = 197
MeV$) for two simulations, one with an instability
($\dot\sigma|_{\tau_0}=-1$)
and one without  ($\dot\sigma|_{\tau_0}=0$).           
.

We notice that for both initial conditions, the system eventually
settles down to the broken symmetry vacuum result as a result of the
expansion. We also considered a radial expansion and obtained similar results \cite{bib:dcc2}. In the radial case, the outstate was reached earlier, but the
number of oscillations where $\chi$ became negative was similar.  
To determine the single particle inclusive pion spectrum we go to 
and adiabatic basis and introduce an interpolating number operator which
interpolates from the initial number operator to the out number operator.
 Introduce mode functions $f^0_k$ which are first order in an adiabatic expansion of the mode equation.
\begin{equation}
f^0_k = {e^{-iy_k(\tau)} \over \sqrt{2 \omega_k}}; ~~ dy_k/dt =
\omega_k,
\end{equation}
\begin{equation}
\phi(\eta,x_\perp,\tau)\equiv{1\over{\tau^{1/2}}} \int \ddk\bigl(\exp(ik
x)
f^0_\kk(\tau)\ a_\kk(\tau)\ + h.c.\bigr)
\end{equation}

 In terms of the initial distribution of particles $n_0(k)$ and $\beta$
we have:
\begin{eqnarray}
n_k(\tau) &\equiv& f(k_{\eta},k_{\perp},\tau) 
= < a^{\dag}_k (\tau) a_k (\tau) >\nonumber\\
&=& n_0(k) + |\beta(k,\tau)|^2 ( 1+2n_0(k)).
\end{eqnarray}
where $\beta(k,\tau) = i (f_k^{0} { \partial f_k \over \partial \tau} -
{\partial f_k^{0 }  \over \partial \tau}f_k)$, 
$n_k(\tau)$ is the  interpolating 
number density. The distribution of
particles is
\begin{equation}
f(k_{\eta}, k_{\perp},\tau) = {d^6 N \over \pi^2 dx_{\perp}^2
dk_{\perp}^2
d\eta dk_{\eta}}.
\label{interp}
\end{equation} 

Changing  variables from $(\eta, k_{\eta})$ to
$(z, y)$ at a fixed $\tau$
we have
\begin{eqnarray}
E{d^3 N \over d^3 k} &=&
{d^3N \over  \pi dy\,dk_{\perp}^2 } = 
\int\pi dz~ dx_{\perp}^2 ~ J ~ f(k_{\eta},
k_{\perp},\tau) \nonumber \\
&& =  A_{\perp} \int dk_{\eta} 
f(k_{\eta}, k_{\perp},\tau)= \int f(k_{\eta}, k_{\perp},\tau) k^{\mu}
d\sigma_{\mu}. \nonumber \\
\end{eqnarray}

To compare our field theory calculation with some standard phenomenological approach, we considered a hydrodynamic calculation with boost invariant kinematics
and determined the spectrum assuming that at hadronization the pions where
at the breakup temperature $T=m_{\pi}$ (as well as $T=1.4 m_{\pi}$), with the distribution given by the 
Cooper-Frye-Schonberg Formula \cite{cfs}
\begin{equation}
E {d^3N \over d^3k} = {d^3N \over \pi dk_{\perp}^2 dy} =
\int g(x,k) k^{\mu} d \sigma_{\mu}
\end{equation}
Here $g(x,k)$ is the single particle relativistic phase space
distribution
function. 
When there is local thermal equilibrium of pions at a comoving
temperature $T_c(\tau)$ one has
\begin{equation}
g(x,k) = { g_{\pi} } \{ {\rm exp} [k^{\mu}u_{\mu}/T_c] -1
\}
^{-1}. 
\end{equation}

The comparison is shown in Figures 2 and 3.

\epsfxsize=7cm
\epsfysize=6cm
\centerline{\epsfbox{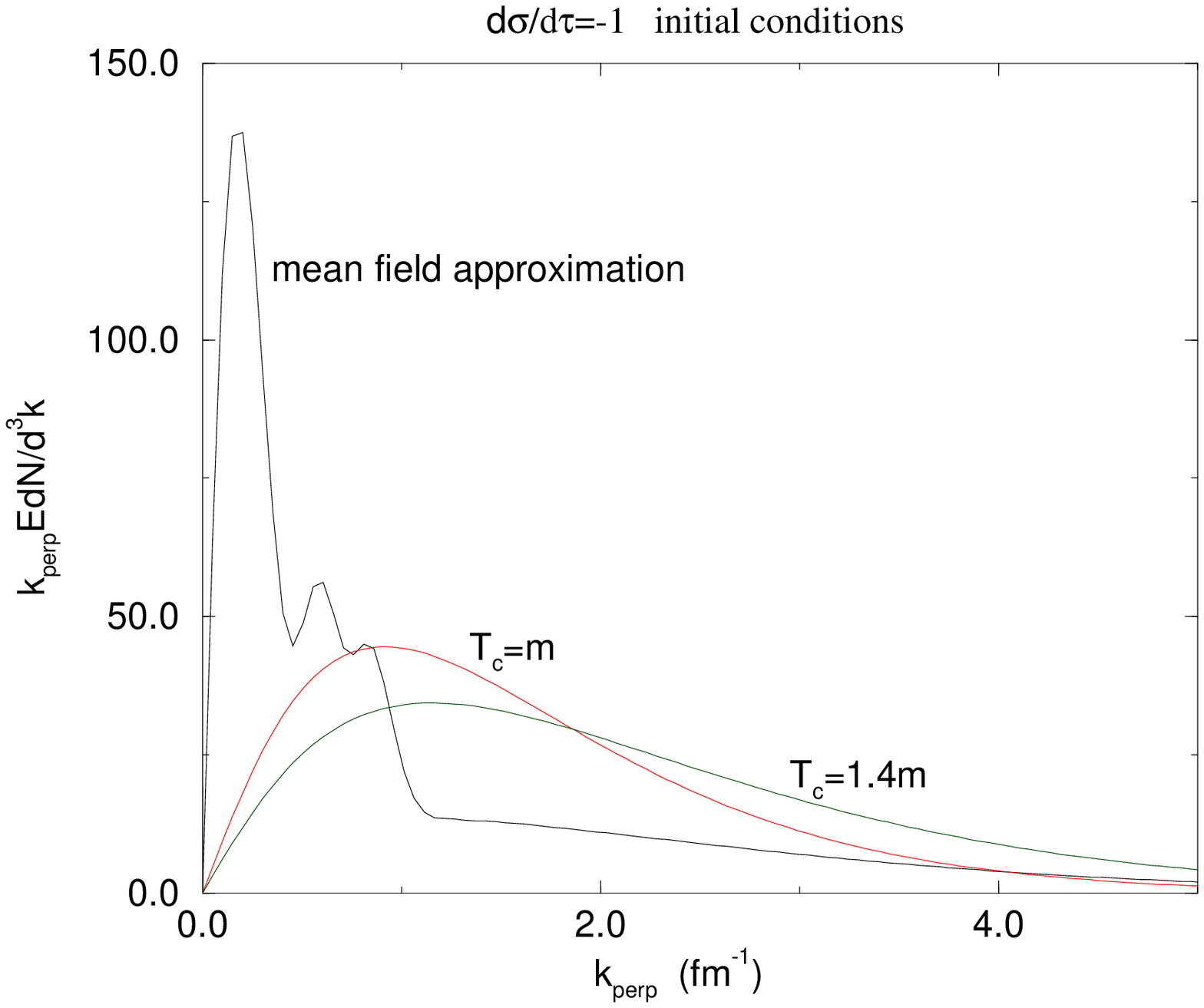}}
\vspace{.35cm}
FIG. 2.Single particle transverse momentum distribution for $\dot{\sigma} =-1$ initial  conditions compared to a local equilibrium
 Hydrodynamical calculation with boost invariance.\\ 
\epsfysize=6cm
\epsfxsize=7cm
\centerline{\epsfbox{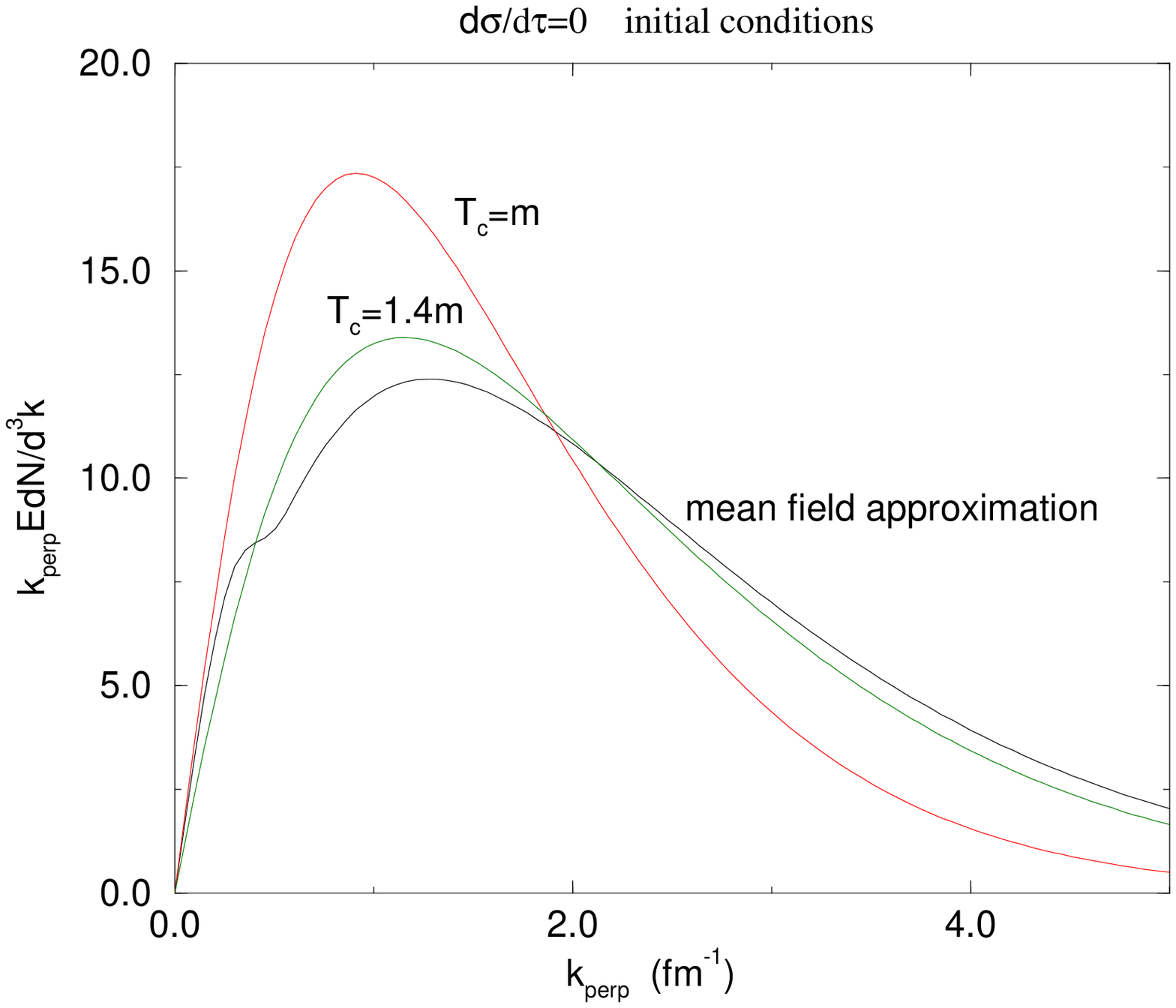}}
FIG. 3. Single particle transverse momentum distribution for
$\dot{\sigma}=0$ initial conditions compared to a local
equilibrium Hydrodynamical calculation with boost invariance.\\          

We therefore find that a non-equilibrium phase transition
taking place during a time evolving quark-gluon or hadronic plasma
can lead to an enhancement of the low momentum 
distribution of pions. 

\subsection{Determination of the Effective Equation of State}

Equation of state is obtained in the frame where the energy
momentum tensor is diagonal-- we are already in that
boost invariant frame :
\[ T_{\mu \nu} = diag \{ \epsilon, p_{\eta}, p_{\perp} \} \]

 When we have massless goldstone pions in the
$\sigma$ model ($H=0$) then $\chi$ goes to zero at large
times.  In the spatially homogenous case:
\[ < T_{00} >= \epsilon  ~~~~ < T_{ij} > = p \delta_{ij} \]
The equation of state becomes $p= \varepsilon/3$ at late times even
though the final particle spectrum is far from thermal
equilibrium.
 
\epsfxsize=10cm
\epsfysize=10cm
\centerline{\epsfbox{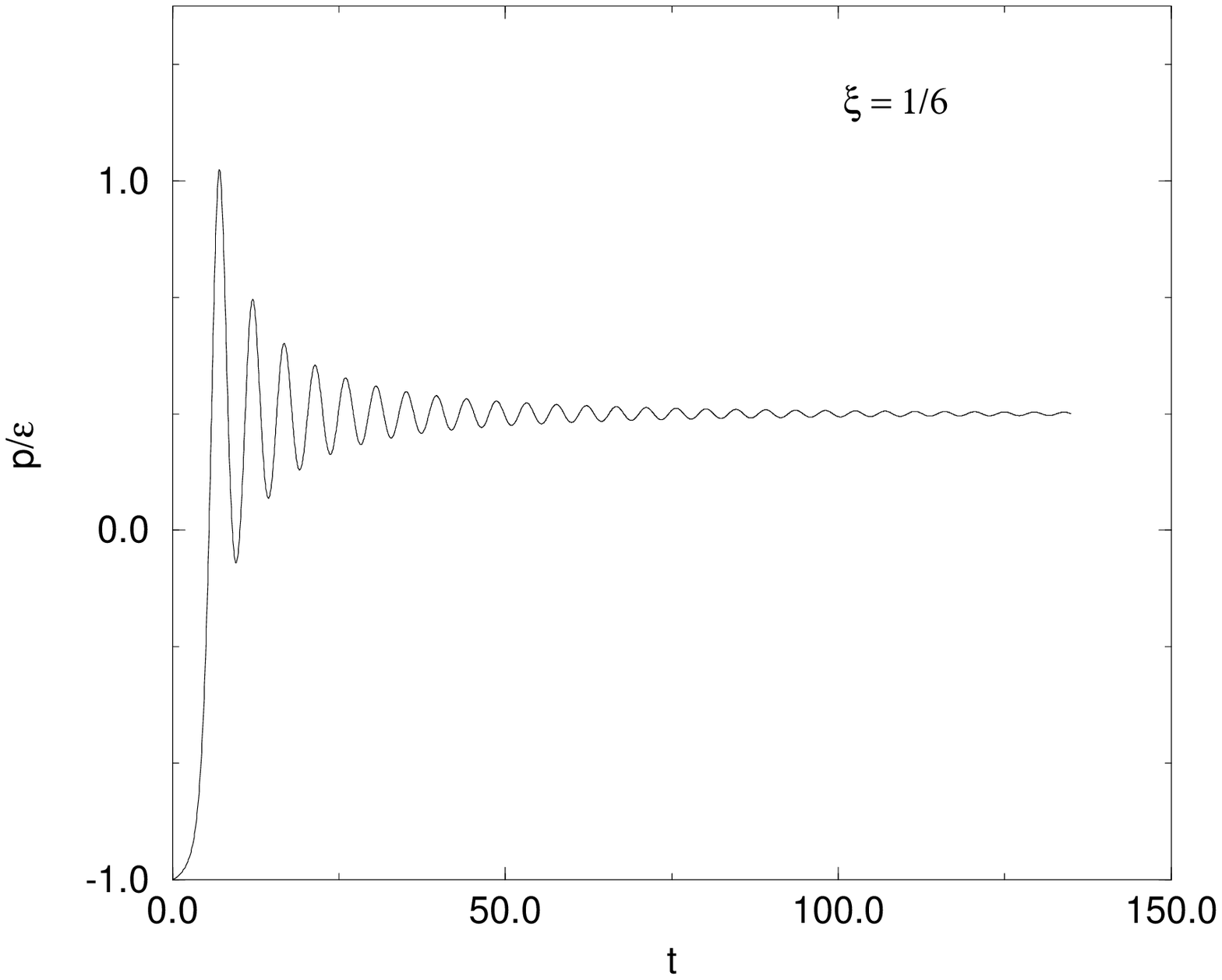}}
\vspace{.2cm}
FIG. 4. Equation of state ${p \over \varepsilon}$ as a function
of $\tau$ for the massless $\sigma$ model where we start from a quench.
\\  

\subsection{Dephasing and looking for  DCC's}
As we have shown in \cite{dephase}, dephasing  justifies the replacement of the
exact Gaussian $\bf \rho$ by its diagonal elements. At large-N or in mean fieldtheory the density matrix is a product of Gaussians in 
 $\phi_k$ space:

\begin{eqnarray}
&&\langle \varphi'_k|{\bf\rho}_{eff} |\varphi_k\rangle = \nonumber\\
&& (2\pi \xi_{\bf k}^2)^{-{1\over 2}}
\exp \biggl\{-{\tilde\sigma_{\bf k}^2\over 8 \xi_{\bf k}^2}(\varphi_k'- 
\varphi_k)^2  
-{1\over 8 \xi_{\bf k}^2}(\varphi_k'+ \varphi_k)^2 \biggr\}~,\nonumber\\
\label{gaueff} 
\end{eqnarray}
\\
After a short while because of dephasing, 
he Gaussian distribution off the diagonal $\varphi_k'=
\varphi_k$ is strongly suppressed
\[ {\xi_k  \over \sigma_k} \approx {\hbar \over 2 k n(k)} << \xi_k \]
This is shown in Fig. 5.
We find no support for
``Schr\"odinger cat'' states in which quantum interference effects
between the two classically allowed macroscopic states at $v$ and $-v$
can be observed.  \\
An ensemble  may be regarded as a classical probability
distribution over classically distinct outcomes 
The particle
creation effects in the time dependent mean field give rise to strong
suppression of quantum interference effects and mediate the quantum to
classical transition of the ensemble.

\vspace{.4cm}
\epsfxsize=7cm
\epsfysize=5cm
\centerline{\epsfbox{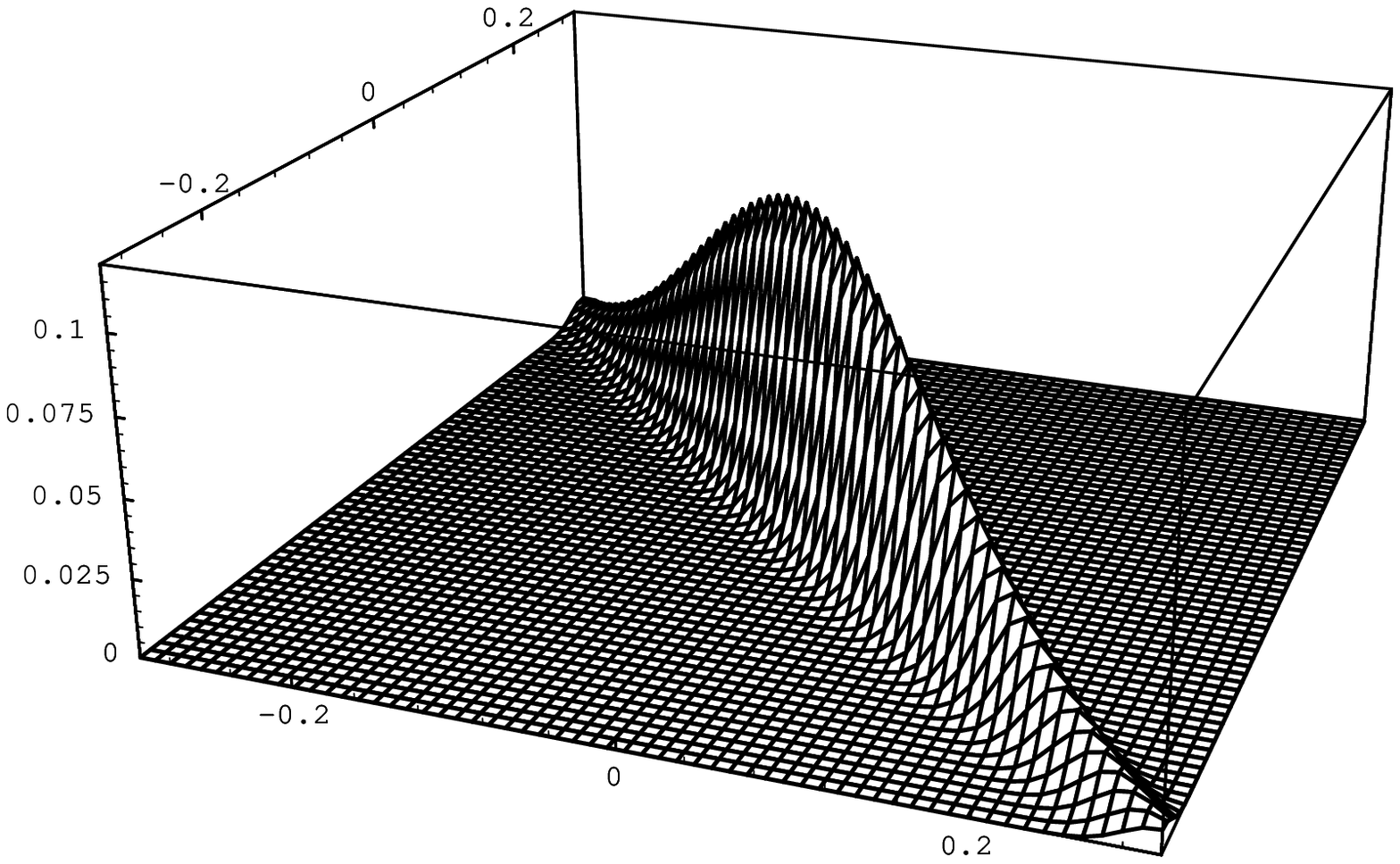}}
\vspace{.35cm}
FIG. 5. The Gaussian $\rho_{eff}$ for $k=.4$ from Ref.\cite{dephase}  illustrating the strong
suppression of off-diagonal components due to dephasing.\\

If  we project the density matrix onto an adiabatic number basis, we can reconstruct
classical field configurations from the diagonal density by replacing
the field operator $a(k)$ by 
\[ a(k) \rightarrow[(n(k)]^{1/2}  e^{i \phi(k)} \]
with $n(k)$ obtained by throwing dice on the density matrix and $\phi$ being randomly
chosen between $ 0 < \phi < 2 \pi$. Typical field configurations as a function of $r$
 (averaging over angles) are shown in Fig. 6. 

\vspace{.4cm}
\epsfxsize=7cm
\epsfysize=5.5cm
\centerline{\epsfbox{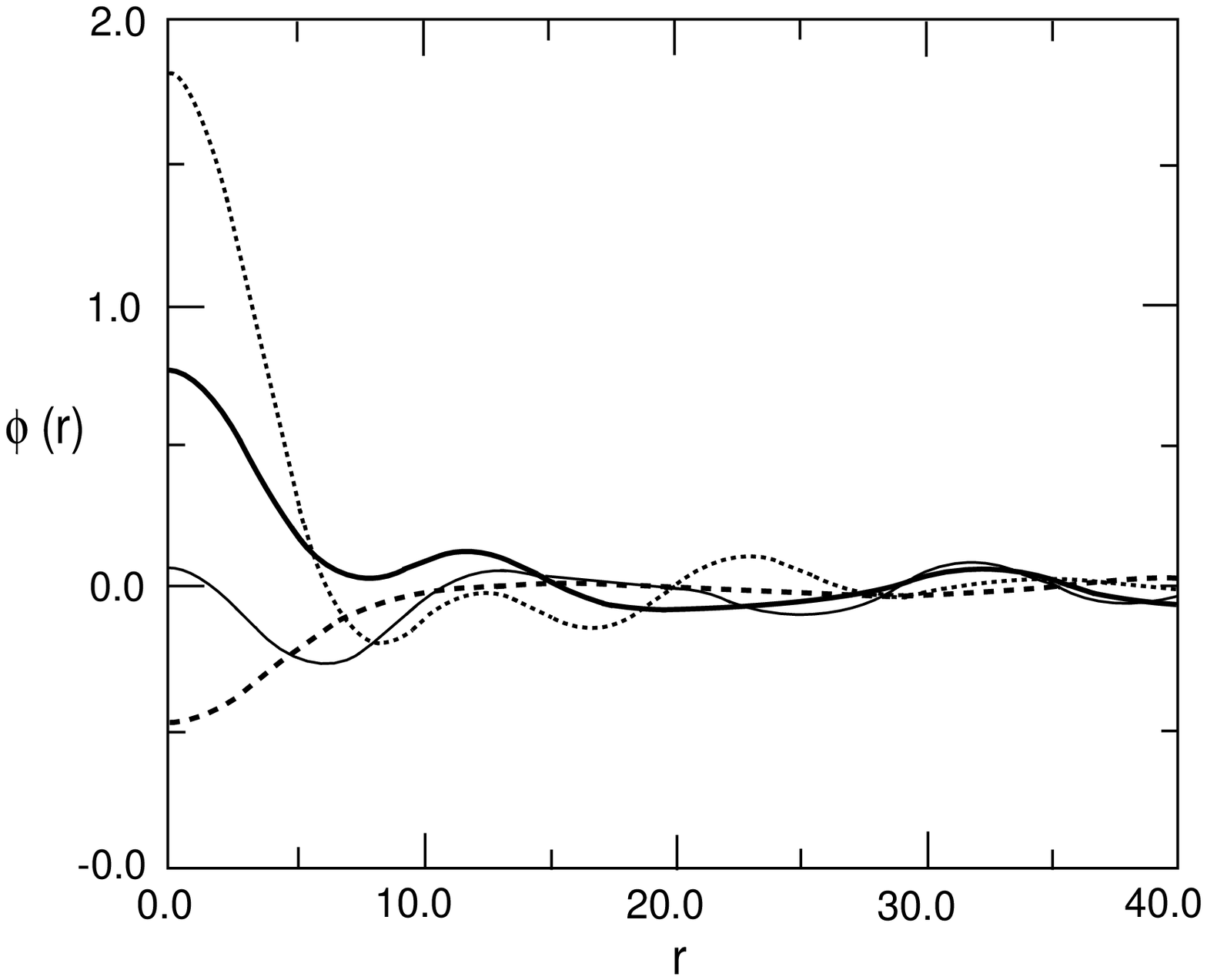}}
\vspace{.35cm}
FIG. 6. Four typical field configurations drawn
from the same classical distribution of probabilities.

\section{INCLUSIVE DILEPTON  PRODUCTION and SCHWINGERS CLOSED TIME PATH FORMALISM }
Schwinger's CTP Formalism  is designed to allow one to calculate  expectation
values of operators in the initial density matrix. One quantity we are interested in for obtaining an effective hydrodynamics is the expectation value of the 
energy momentum tensor:
\begin{equation}
< in | T^{\mu \nu} (x) |in >  \equiv (\epsilon + p) u^{\mu} u^{\nu}
 - p g^{\mu \nu} 
\end{equation}
where $T^{\mu \nu} (x)$ is the Field Theory energy momentum tensor.
Also by fourier transforming this energy momentum tensor and looking in a comoving frame, we can ask how much energy is in the ``free'' part of various
components and define an equivalent number of quanta by dividing by 
$ \hbar \omega_k$ for each specie.

If we consider the inclusive production of electron positron pairs
the probablility amplitude is 
\begin{eqnarray}
  <e^-(k,s) e^+(k',s')   X | i > = < X |b^{(out)}_{k,s}  d^{(out)}_{k',s'} |i>.
\nonumber 
\end{eqnarray} 
The  inclusive distribution function for  dileptons : 
\begin{eqnarray}
{E_k \over m} {E_k'\over m} {d^6N \over [d^3k] [d^3k']} 
\equiv   < i|d^
{+(out)}_{k',s'}  b^{+(out)}_{k,s} b^{(out)}_{k,s}
  d^{(out)}_{k',s'}|i>.\nonumber
\end{eqnarray}
Using the relations between $b$, $d$ to $\Psi$ and the Free ``out''  fields we obtain
\begin{eqnarray}
&&   _{in} < i|\int d^3 x_1 d^3 x_2 d^3 x_3 d^3 x_4 \nonumber \\
&& e^{ik(x_2-x_4)}\{ u^+_{k,s}\Psi^{(out)}(x_2) \}\{\Psi^{(out)+}(x_4) u_{k,s} \}\nonumber\\
&& e^{ik'(x_1-x_3)} \{ v^+_{k',s'} \Psi^{(out)}(x_3) \} \{ \Psi^{(out+)}(x_1) v_{k',s'} \}|i> \nonumber
\end{eqnarray}
Now using the weak asymptotic condition \cite{ref:LSZ} that 
\begin{equation}
\Psi |_{t \rightarrow \infty} =  Z^{1/2} \Psi^{(out)}
\end{equation}
inside of matrix elements as well as the equation of motion of the spinors
and the identity: 
\begin{equation}
\int_{t_1}^{t_2}  {dF \over dt} = F(t_2) - F(t_1), \label{eq:iden}
\end{equation} 
we obtain:
 
\begin{eqnarray}
 && _{out} <e^-(k,s)~ e^+(k',s') ~ X | P_1 P_2 >_ {in} = \nonumber \\
&& i^2 Z^{-1} \int d^4x_1 d^4 x_2
e^{i(k x_2 + k'x_1)}   \nonumber \\
&& \bar{u}_{k,s}{\cal{D}}~~ \, _{out} < X| {\cal{T}} \{\Psi(x_2) {\bar \Psi}(x_1) \}|P_1 P_2 >_ {in}  \bar{{\cal{D}}} v_{k',s'}
\end{eqnarray}
Squaring this amplitude and summing over $X$ we obtain:
\begin{eqnarray}
&&{E_k \over m} {E_k'\over m} {d^6N \over [d^3k] [d^3k']}(k,k';s,s')=  \nonumber \\
&&\int d^4 x_1
d^4 x_2 d^4 x_3 d^4 x_4  e^{ik(x_2-x_4)}
 e^{ik'(x_1-x_3)} 
\bar {v}_{k',s'}{\cal{D}}_{x_3} \bar{u} _{k,s}{\cal{D}}_{x_2} \times \nonumber \\
&& \,  _{in} < P_1 P_2 |
 {\cal{T}}^* \{  \Psi(x_3) \bar{\Psi}(x_4) \}  {\cal{T}} \{
\Psi(x_2) \bar{\Psi}(x_1)  \}| P_1 P_2 >_ {in} \times \nonumber \\
&& \bar{ \cal{D}}_{x_4}u_{k,s} 
\bar{\cal{D}}_{x_1}v_{k',s'}   \nonumber \\
\label{eq:dilep2}
\end{eqnarray}
The matrix element involved here,
\begin{equation}
  _{in} < P_1 P_2 |
 {\cal{T}}^* \{  \Psi(x_3) \bar{\Psi}(x_4) \}  {\cal{T}} \{
\Psi(x_2) \bar{\Psi}(x_1)  \}| P_1 P_2 >_ {in} \label{eq:seectp} 
\end{equation}
is precisely the type of Green's function that is obtained from the
generating functional of Schwinger's CTP formalism

The Lagrangian we will use to determine this 4 point function is
the $O(4)$ linear $\sigma$ model +  Electrodynamics.

This Lagrangian has 3 pieces:
The mesons  form an $O(4)$ vector $\Phi = ( \pi_i,\sigma)$.  This strongly
interacting Lagrangian is given by
\begin{equation}
L_{strong} = -{ 1 \over 2} \phi_i (\Box + \chi) \phi_i + {\chi^2 \over 4 \lambda}
+  {1 \over 2} \chi v^2 + H \sigma
\end{equation}
$$\chi = \lambda (\Phi \cdot \Phi-v^2).$$

To this we add the free lepton and Photon Lagrangian:
\begin{equation}
L_0 = - {1 \over 4} F_{\mu \nu} F^{\mu \nu} - {1 \over 2 \alpha} (\partial \cdot A)^2
+ \bar{\Psi} [ i\gamma_{\mu} \partial^{\mu}-m ] \Psi
\end{equation}  

The interaction of the Photons with the pion plasma and the leptons is given by
\begin{eqnarray}
L_{int}[\phi_i,A_{\mu},\Psi, \bar{\Psi}] &&= {e^2 \over 2} (\phi_1^2+\phi_2^2) A_{\mu} A^{\mu} \nonumber \\
&&+ 
e(\phi_1 \partial_{\mu} \phi_2- \phi_2 \partial_{\mu} \phi_1) A^{\mu} \nonumber \\
 -e \bar{\Psi} \gamma^{\mu} \Psi A_{\mu}+ {\cal L_A}.\label{eq:int}
\end{eqnarray}

If we treat the electromagnetic interactions perturbatively in $e^2$
and the pions in the mean field approximation we obtain the
graph shown in the Figure 7
\epsfxsize=5.5cm
\epsfysize=4cm
\vspace{1cm}
\centerline{\epsfbox{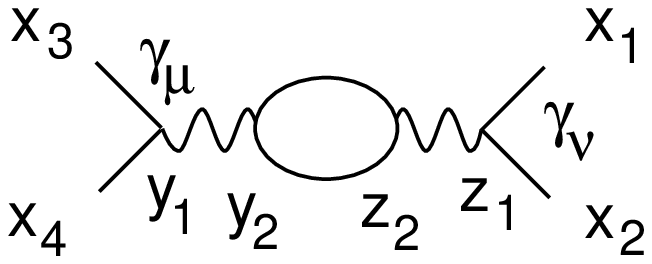}}
\vspace{.2cm}

FIG. 7. Leading contribution from the plasma to dilepton production.The 4 fermion Graph is to be evluated using the Matrix CTP Green's functions.
\\

The inverse propagators in the LSZ representation lop off the external
legs and put the leptons on mass shell. One is left with:
\begin{equation}
{E_k \over m} {E_{k'} \over m} {dN \over d^3k d^3k'} = M_{\mu \nu}(k,s; k's')
W^{\mu\nu }(k,k') \end{equation}
where

\[
M^{\mu \nu}(k,s; k's') = {\bar v}(k',s') \gamma^{\mu}u(k,s) {\bar u}(k,s) \gamma^{\nu} v(k',s')
\]

\begin{eqnarray}
&&W_{\mu \nu} (k,k')\equiv W_{\mu \nu}^1+W_{\mu \nu}^2+W_{\mu \nu}^3  \nonumber\\
&&=i e^4 \int d^4 y_1 \int d^4 y_2 \int d^4 z_1 \int d^4 z_2
e^{i (k+k') (z_1-y_1)}   \nonumber \\
&&\times[\, D_{\mu \sigma}^{ret} (y_1,y_2) \Pi^{\sigma \lambda 21} (y_2,z_2) D_{\lambda \nu}^{ adv}(z_2,z_1) \nonumber \\ 
&&+ D_{\mu \sigma}^{21} (y_1,y_2) \Pi^{\sigma \lambda}_{adv} (y_2,z_2) D_{\lambda \nu}^{adv}
(z_2,z_1) \nonumber \\
&&+ D_{\mu \sigma}^{ret} (y_1,y_2) \Pi^{\sigma \lambda}_{ret} (y_2,z_2)
 D_{\lambda \nu}^{21} (z_2,z_1)] \nonumber
\end{eqnarray}

If we want the invariant Mass distribution function when 
\[ M^2 = q^2; ~~ q=k+k'\]
we obtain
\begin{eqnarray}
{dN \over d^4 q}&&= 2{ q^0 dN \over dM^2 d^3q}  = R_{\mu \nu} (q)  W^{\mu
\nu}(q). \label{eq:dil} \end{eqnarray}

where

\begin{eqnarray}
R_{\mu\nu}(q) &&\equiv \int {[d^3k] \over E_k}  {[d^3k'] \over E_{k'}}\delta^4
(q-k-k'){\bar L}_{\mu\nu}(k,k') \nonumber \\
&& ={1 \over (2 \pi)^6} { 2 \pi \over 3}  (1-{4 m^2 \over q^2})^{1/2}(1+{2m^2
\over q^2})(q^{\mu} q^{\nu}- g^{\mu \nu} q^2) \nonumber \\ 
\end{eqnarray}

If we were doing an ordinary perturbation theory calculation analytically
we could use the translational invariance of the polarization tensor:
\begin{equation}
\Pi^>_{\mu \nu}(y_2,z_2) = \int [d^4q] e^{-iq (y_2-z_2)} \Pi^>_{\mu \nu}(q)
\end{equation}
and the representation of the free Photon propagator in Feynman Gauge:
\begin{equation}
D_{F\mu \nu}(z_2,z_1) = \int [d^4k]	 e^{-ik (z_2-z_1)}{ g_{\mu \nu} \over k^2+i
\epsilon}  
\end{equation}
to obtain:

\begin{eqnarray}
W_{\mu \nu} (k,k')&& = -i e^4 {(2 \pi)^4 \delta^4(0) \over q^4}\Pi_{\mu
\nu}^>(q^2)\nonumber \\
&& = -i e^4 VT \int {d^4q \over q^4} \delta^4(q-k-k') \Pi_{\mu \nu}(q^2)
\label{eq:whom} 
\end{eqnarray}
The other terms have the photon on mass shell so they
give no contribution to the particle production rates. 

In this homogeneous case we therefore obtain the usual result \cite{ref:mclerran} \cite{ref:ruuskanen}:
\begin{equation}
 {q^0 dN \over dM^2 d^3q VT} ={e^4 \over (2 \pi)^6}{ i \pi \over 3}
 {\Pi^{> \mu}_{\mu}(q^2) \over q^2} (1-{4 m^2 \over q^2})^{1/2}(1+{2m^2 \over
q^2}), \end{equation}

Let us first look at the case of a thermal plasma where we can calulate
everything analytically. The vacuum polarization graph can be found using 
the following expansion of the pion field to calculate the Finite Temperature
pion Green's functions:
\begin{equation}
\phi(x,t) = \int {[d^3k] \over 2 \omega_k}\left [  \exp (i k x)  a_k + \exp (-i k
x)  b^{\dag}_k \right] \end{equation}
The creation and annihilation operators obey the commutation relations:
\begin{equation}
[a_k, a^{\dag}_{k'}] = [b_k,b^{\dag}_{k'}] = (2 \pi)^3 \delta^3 (k-k')
\label{eq:com1} \end{equation}
And the phase space number densities $n_k^+$ and $n_k^-$ are defined by
\begin{eqnarray}
 \langle a^{\dag}_k a_{k'} \rangle &&= (2 \pi)^3 n_k^+ \delta^3(k-k') \nonumber \\
 \langle b^{\dag}_k b_{k'} \rangle &&= (2 \pi)^3 n_k^- \delta^3(k-k')
\label{eq:ndef} \end{eqnarray}
so that the total number of positively charged particles is given by:
\begin{equation}
N^+ = \int d^3k  \langle a^{\dag}_k a_k \rangle = \int d^3x d^3k n^+_k
\end{equation}
For the case of neutral plasma in thermal equilibrium at inverse temperature
$\beta$ we have
\[ n_k^+ = n_k^- =1/(e^{ \beta \omega}-1)
 \]
Using the fact that for a free pion gas:
 \begin{eqnarray}
   &&\langle\phi^\dagger(x) \phi(y)\rangle_{th}\nonumber\\
&&=
   \int {d^3k \over 2\omega_k (2\pi)^3} [ n^+_k e^{ik\cdot(x-y)}
   +(1+n^-_k) e^{-ik\cdot(x-y)}]\ , \nonumber \\
   &&\langle\phi(x) \phi^\dagger(y)\rangle_{th}\nonumber\\
&&=
   \int{d^3k \over 2\omega_k (2\pi)^3}[ n^-_k e^{ik\cdot(x-y)}
   +(1+n^+_k) e^{-ik\cdot(x-y)}]\ , \label{eq:thprop}
 \end{eqnarray}

one finds that the vacuum polarization tensor is given by
\begin{eqnarray}
&&-i \Pi_{\mu\nu}^>(q^2)=\int {d^3k_1 \over 2\omega_1(2\pi)^3}
   {d^3k_2 \over 2\omega_2(2\pi)^3} (2\pi)^4  \nonumber\\
&&\left\{ 
   (k_1-k_2)_\mu (k_1-k_2)_\nu \right. \nonumber \\ 
 &&  [n^-_{k_1}n^+_{k_2} \delta^4(q-k_1-k_2)  \nonumber\\
&&  +
   (1+n^-_{k_1})(1+n^+_{k_2})\delta^4(q+k_1+k_2)] \nonumber \\
   && (k_1+k_2)_\mu(k_1+k_2)_\nu  \nonumber\\
&& \left.[ n^-_{k_1}(1+n^-_{k_2}) 
   +n^+_{k_1}(1+n^+_{k_2})]\delta^4(q-k_1+k_2) \right\}, \nonumber \\ 
 \label{Wth}:
\end{eqnarray}
 The three terms in the above relation correspond to pion pair
 annihilation, creation and bremsstrahlung, respectively.
The delta functions show that only the annihilation process survives for
$M^2$
above the dilepton threshold.
We therefore obtain:
\begin{equation}
-i \Pi_{\mu}^{\mu >}(q) = q^2  (1-{4m_\pi^2 \over q^2}) 
\int_{\omega^-}^{\omega^+} {d\omega \over q} n^+_k n^-_{q-k} \;\; ,
\end{equation}
 with $\omega^\pm=(q_0 \pm q \sqrt{1-4m_\pi^2/M^2})/2$ and 
 $q_0 = \sqrt{M^2 + q^2}$. 
Thus the the dilepton production rate from a homogenous thermal plasma of pions
is  given by: 
\begin{equation}
  {1 \over V T}{dN_{\ell^+\ell^-}^{(2)}\over dM d^3q}
   ={\alpha^2  B \over 48 \pi^4} {M \over q_0} (1-{4m_\pi^2 \over M^2})
   \int_{\omega^-}^{\omega^+} {d\omega \over q} n^+_k n^-_{q-k} .
   \label{eq:theory}
 \end{equation}

\subsection{Finite Time Effects}

For our numerical simulation of the interacting plasma we can
only follow the time evolution of the plasma for a fixed time $2T$ which
is typically about 100 fermis. 

Therefore  we need to make sure we have the causal formulation so that
dileptons at time $T$ only get contributions from processes occuring at times $ t < T$. So we
should not use Feynman propagators but instead use the CTP matrix propagators
when we investigate the theory at finite times. 
Also the interacting plasma is not time translationally invariant so we
must use a non-covariant formalism.
In the mean field approximation (as well as for the pion gas) there is factorization of four
point functions so that the current-current  correlation function  takes the form
\begin{eqnarray}
&&<J^\mu(x)J^{\dagger \nu}(y)>\nonumber\\
&&=<\Phi^\dagger(x)\Phi(y)><\partial^\mu\Phi(x)\partial^\nu \Phi^\dagger(y)> 
\nonumber\\
&&-<\Phi^\dagger(x) \partial^\nu\Phi(y)><\partial^\mu\Phi(x)\Phi^\dagger(y)>
\nonumber \\ 
&&- <\partial^\mu\Phi^\dagger(x) \Phi(y)> <\Phi(x)\partial^\nu
\Phi^\dagger(y)>\nonumber\\ &&+ <\partial^\mu\Phi^\dagger(x)
\partial^\nu\Phi(y)> <\Phi(x) \Phi^\dagger(y)>. \end{eqnarray}

If we insert the mode
expansion of the charged pion fields:
\begin{equation}
\Phi(x,t) = \int [d^3k] \left [  \exp (i k x) f_k(t) a_k + \exp (-i k x) f^*_k (t) b^{\dag}_k \right]
\end{equation}
and define
phase space number densities $N_k^+$ and $N_k^-$   by
\begin{eqnarray}
 \langle a^{\dag}_k a_{k'} \rangle &&= (2 \pi)^3 N_k^+ \delta^3(k-k') \nonumber \\
 \langle b^{\dag}_k b_{k'} \rangle &&= (2 \pi)^3 N_k^- \delta^3(k-k')
\end{eqnarray}

we obtain:
\begin{eqnarray}
&&<J_\mu(x)J_\nu(y)>=\nonumber \\
&& e^2 \int {d^3k\over (2\pi)^3}{d^3p\over (2\pi)^3}
e^{-i(\vec{k}-\vec{p})\cdot(\vec{x}-\vec{y})}{1 \over i} G_{\mu
\nu}(\vec{k},\vec{p};t_x,t_y), \label{Jcorrelator} \nonumber \\
\end{eqnarray}
where
\begin{eqnarray}
{1 \over i}{G_{\mu \nu}(\vec{k},\vec{p};t_x,t_y) }&&=
A(\vec{k}) K^+_{\mu \nu}(\vec{p}) + K^-_{\mu \nu}(\vec{k})A(\vec{p}) \nonumber \\
&&-N^{+\nu}(\vec{k}) M^{+\mu}(\vec{p}) - {M}^{-\mu}(\vec{k}) {N}^{-\nu}(\vec{p}),
\nonumber \\
\end{eqnarray}

and
\begin{equation}
\begin{array}{lr}
K_{\mu \nu}^{\pm}(\vec{k})  \equiv \left \{ \begin{array}{cccc}
  B(\vec{k}) &\mu =0 & \nu=0  &  \\
 \pm i k_jD(\vec{k})  & \mu=0  &\nu=j &   \\
 -(\pm)i k_i C(\vec{k})  & \mu=i &\nu=0 &  \\
  k_ik_j A(\vec{k})   &\mu=i  & \nu=j &   
\end{array} \right \} ,           
\end{array}
\label{pi}
\end{equation}
\begin{equation}
\begin{array}{lr}
M^{\pm}_\mu(\vec{k})  \equiv \left \{ \begin{array}{cc}
  D(\vec{k})  &\mu =0   \\
  -(\pm)i k_i A(\vec{k})  & \mu=i 
 \end{array} \right \} ,           
\end{array}
\label{M}
\end{equation}

\begin{equation}
\begin{array}{lr}
N_{\nu}^\pm (\vec{k})  \equiv \left \{ \begin{array}{cc}
  C(\vec{k})  &\nu =0   \\
  -(\pm)i k_j A(\vec{k})  & \nu=j 
 \end{array} \right \} ,           
\end{array}
\label{N}
\end{equation}
and

\begin{eqnarray}
&&A(\vec{k};t_x,t_y)=(1+N_k)[f^\ast_k(t_x)f_k(t_y)]^\ast 
+       N_k f^\ast_k(t_x)f_k(t_y)\nonumber\\
&&B(\vec{k};t_x,t_y)=(1+N_k)[\dot{f}^\ast_k(t_x)\dot{f}_k(t_y)]^\ast 
+       N_k \dot{f}^\ast_k(t_x)\dot{f}_k(t_y)\nonumber\\
&&C(\vec{k};t_x,t_y)=(1+N_k)[f^\ast_k(t_x)\dot{f}_k(t_y)]^\ast 
+       N_k f^\ast_k(t_x)\dot{f}_k(t_y)\nonumber\\
&&D(\vec{k};t_x,t_y)=(1+N_k)[\dot{f}^\ast_k(t_x)f_k(t_y)]^\ast 
+       N_k \dot{f}^\ast_k(t_x)f_k(t_y). \nonumber \\
\end{eqnarray}

Contracting $G_{\mu \nu}(\vec{k}, \vec{p})$ with  $\tilde{L}^{\mu \nu} =
q^\mu q^\nu - g^{\mu \nu} q^2$ we obtain:

\begin{eqnarray}
&&\tilde{L}^{\mu \nu}G_{\mu \nu}(\vec{k} \vec{p};t_x,t_y) = \nonumber \\
&&[(\vec{q} \cdot (\vec{p}+\vec{k}))^2
+ (q_0^2-\vec{q} \cdot \vec{q}) (\vec{p}+\vec{k})\cdot (\vec{p}+\vec{k}) ]
 A(\vec{p}) A(\vec{k}) \nonumber \\
&&- i q_0 (\vec{p}+\vec{k})\cdot
\vec{q}[A(\vec{k})D(\vec{p})-D(\vec{k})A(\vec{p}) \nonumber \\
&&-A(\vec{k}) C(\vec{p})
+C(\vec{k}) A(\vec{p})] \nonumber \\
&&+ \vec{q}\cdot \vec{q}[A(\vec{k})B(\vec{p})+B(\vec{k})A(\vec{p})-C(\vec{k})
D(\vec{p}) - D(\vec{k}) C(\vec{p})] \nonumber \\   
\end{eqnarray}

where 
\[ \vec{q} = \vec{k} - \vec{p}
\]
 At the special case where $\vec{q}=0$ we obtain:
\begin{equation}
\tilde{L}^{\mu \nu}G_{\mu \nu}(\vec{k} \vec{k};t_x,t_y)  = 4 q_0^2 \vec{k}\cdot \vec{k}
  A(\vec{k}) A(\vec{k}) 
\end{equation} 
Here $ \vec{k}\cdot \vec{k} \rightarrow {q_0^2 - 4 m^2 \over 4}$ in the infinite time limit. 

\subsection{pion gas}
To see what the effects of finite $T$ might be, let us look at the
case where everything is analytically known, namely the pion gas
we discussed before. So we use the known values of $A,B,C,D$ appropriate to the pion
gas where 
\[ f_k(t) = { e^{-i \omega_k t}  \over \sqrt{2 \omega_k}}. \]

First let us look at the effect of just putting a finite time cutoff 
into the McLerran formula for $W$.
Using the covariant form of the photon propagator and just cutting off the
internal integrations to run from $-T$ to $T$ one  obtains:
\begin{eqnarray}
 W_{\mu \nu}^{cutoff} (k, k')&& = i{e^4 \over q^4} \int d^3y_2 
\int d^3 z_2 \int_{-T}^T dy_{2_0} \int_{-T}^T dz_{2_0} \times \nonumber \\
&& e^{-iq(y_2-z_2)} \Pi^{\mu \nu} (y_2,z_2) \nonumber \\
&&= i V {e^4 \over q^4} \int_{-T}^T dy_{2_0} \int_{-T}^T dz_{2_0} 
e^{-iq_0(y_{2_0}-z_{2_0})} \times \nonumber\\
&&\int [dK] G^{\mu \nu} (\vec{K},\vec{K}-\vec{q}; y_{2_0},z_{2_0})
\end{eqnarray}
Again looking only at the place  $\vec{q}=0$ and keeping only the pion
annihilation contribution:
\begin{equation}
\tilde{L}^{\mu \nu}G_{\mu \nu}(\vec{k} \vec{k};y_{2_0},z_{2_0})  =
 {n_k^2 \over \omega_k^2}  q_0^2 \vec{k}\cdot \vec{k}  e^{2 i
\omega_k (y_{2_0}-z_{2_0})}
\end{equation}
we can perform   
perform the integration over $T$ and obtain the factor:
\[
4 ({\sin[q_0-2 \omega_k] T \over q_0-2 \omega_k})^2 \]
which in the limit $ T \rightarrow \infty$ becomes the factor
\[ 2 T \times 2 \pi \delta( q_0- 2 \omega_k) 
\]
Thus we see that in the infinite time limit, the back to back dileptons
get contributions only from pion pairs in the plasma with zero combined
three momentum, each carrying energy $= q_0/2$. In what follows we
want to see how doing a causal calculation changes the way in which
we go on mass-shell from this simple replacement of the delta function
by a representation of the delta function.  We will find extra terms
which only slowly go to zero with the time $T$.

We now use the 3 dimensional form for the propagators:
\begin{eqnarray}
&& D_{\mu \nu}^{ret}(x,y) = i \Theta (x_{0}-y_{0}) g_{\mu \nu}
e^{-\varepsilon (x_0-y_0)} \times \nonumber \\ 
&&\int { [dq] \over 2 \bar{q}_0} e^{i
\vec{q} \cdot (\vec{x}-\vec{y})}[e^{-i\bar{q}_0(x_0-y_0)}-
e^{i\bar{q}_0(x_0-y_0)}] \nonumber \\
 && \equiv  i \Theta (x_{0}-y_{0})e^{-\varepsilon (x_0-y_0)} g_{\mu
\nu} \int  [dq]  e^{i \vec{q} \cdot (\vec{x}-\vec{y})}  \triangle
[\bar{q}_0(x_0-y_0)] \nonumber \\
\end{eqnarray} 

In our numerical simulations we assumed spatial homogeneity so that
the vacuum polarization has the form:
\begin{eqnarray}
\Pi^{> \mu \nu} (xy) =&& i \langle  J^{\mu}(x) J^{\nu}(y) \rangle \nonumber \\
=&&\int [dK] \int [dP] e^{-i (\vec{K}-\vec{P}) \cdot (\vec{x}-\vec{y})} 
G^{\mu \nu} [ \vec{K}, \vec{P};x_0,y_0] \label{eq:pimunu} \nonumber \\
\end{eqnarray}
Inserting these into the expression  for $W^{\mu \nu}(k,k')$ we then obtain

\begin{eqnarray} 
&&W^{\mu \nu (1)}(k,k') = i e^4  V \int_{-T}^T  dy_{10}\int_{-T}^T  dz_{10}
\int_{-T}^{y_{10}}  dy_{20} \int_{-T}^{z_{10}} dz_{20}  \nonumber\\
&&\int [dK] e^{- \varepsilon (y_{10}-y_{20})}e^{- \varepsilon (z_{10}-z_{20})}
 e^{i q_0 (z_{10}-y_{10})}
\nonumber\\
&& \triangle [ |\vec{q}|(y_{10}-y_{20})] G^{\mu \nu} (\vec{K},\vec{K}-\vec{q}; y_{20},z_{20})
 \triangle [ |\vec{q}|(z_{20}-z_{10})] \label{eq:wk1} \nonumber \\
\end{eqnarray}
where 

\[ \vec{q} = \vec{k} + \vec{k'} \]
It is this contribution which persists when the time cutoff  $ T \rightarrow \infty$.

For the other two contributions we obtain:
\begin{eqnarray} 
&&W^{\mu \nu (2+3)}(k,k') = - 2 e^4  V \nonumber \\
&& \int_{-T}^T  dy_{10}\int_{-T}^T  dz_{10}
\int_{-T}^{z_{10}}  dz_{20} \int_{-T}^{z_{20}}  dy_{20} \int [dK] 
Im [ e^{i q_0 (z_{10}-y_{10})} \nonumber \\
&&[G^{\mu \nu} (\vec{K},\vec{K}-\vec{q}; y_{20},z_{20})- G^{\mu \nu} (-\vec{K},\vec{q}-\vec{K}; z_{20},y_{20})] \nonumber \\
&& \times {e^{- i |\vec{q}|(y_{10}-y_{20})} \over 2|\vec{q}|} \triangle
 [ |\vec{q}|(z_{20}-y_{10})] ] \label{eq:wk23}
\end{eqnarray}
Things simplify dramatically at the place where  $\vec{q} =0$. At that point
the second and third contributions vanish and we have
\begin{eqnarray}
&&{1 \over VT} {dN \over d^{n+1} q} = {4 q_0^2  \over (2 \pi)^{2n}} {2  B \pi
\over 3}
\Omega_n \int k^{n+1} dk  ~~
F[k , \vec{q} =0, q_0] \nonumber \\
&&F[k , \vec{q} =0, q_0] = i{ e^4  \over T}
\int_{-T}^T  dy_{10}\int_{-T}^T  dz_{10}
\int_{-T}^{y_{10}}  dy_{20} \int_{-T}^{z_{10}}  dz_{20} ~~ ~  \nonumber \\
&& e^{i q_0~(z_{10}-y_{10})}\triangle_0(y_{10}-y_{20})[A(k ,y_{20},z_{20})]^2
\triangle_0 (z_{20}-z_{10})   \nonumber \\ \end{eqnarray}
where
\[   A(k ,t,t') = (1+n_k){ e^{-i \omega_k (t-t')}\over 2 \omega_k} +n_k
{ e^{i \omega_k (t-t')}\over 2 \omega_k} \]
and
\[  \Delta_0(t-t') = - i {\sin m_{\gamma}(t-t') \over m_{\gamma}};~~ \Omega_n=
2{\pi^{n/2} \over \Gamma(n/2)} \]
where $n$ is the number of spatial dimensions.

In the limit $m_{\gamma} \rightarrow 0$ we have :
\[  \Delta_0(t-t')= -i (t-t') e^{ -\epsilon \vert t-t' \vert} \]

The result for the annihilation part for massless photons is:
\begin{eqnarray}
 F_{ann} &&= -{e^4  N_k^2(k^2) \over 64 q_0^4\omega_k^4  (q_0- 2 \omega_k )^2
T} \times
\nonumber \\
&& [ 16 \omega_k^4 +q_0^4 - 8
\omega_k^2 q_0^2 \cos [2 (2\omega_k-q_0) T]
\end{eqnarray}
with
\[ \omega_k^2 = k^2 + m^2 \]
We can rewrite this as:
\begin{eqnarray}
F_{ann} &&=
 -{e^4  N_k^2 \over 4 q_0^2 \omega_k^2} \times \nonumber \\
&&
\{ {\sin^2 [(2 \omega_k-q_0) T] \over (q_0- 2 \omega_k)^2 T}  +
{(q_0+2 \omega_k)^2 \over 16 q_0^2 \omega_k^2 T} \}
\end{eqnarray}

We would like to compare the representation  of the 
delta function squared found here (renormalized
to one at the delta function) to the result of just naively putting
a cutoff into the covariant calculation which gave:

\begin{equation}
{ \sin^2( \{q_0-2  \omega_k\} T) \over  \{q_0-2 \omega_k\}^2 T^2}
\end{equation}
The CTP formalism which preserves causality instead gives:
\[ \{ {\sin^2 [(2 \omega_k-q_0) T] \over (q_0- 2 \omega_k)^2 T^2}  +
{(q_0+2 \omega_k)^2 \over 16 q_0^2 \omega_k^2 T^2} \}
\]
The last term makes a very small difference at large $T$.

In obtaining this result we assumed that
\[  \varepsilon T \rightarrow \infty  ~~{\rm as}~~ \varepsilon \rightarrow 0. \]

For the brehmstrahlung contribution, we find
\begin{eqnarray}
&&F_{brehm}=-4 {e^4 n_k (1+n_k) \over \pi T  \left( {{{m_{\gamma}}}^2} +
{{\varepsilon }^2} \right) ^2 \omega_k^2 (q_0^2 -m_{\gamma}^2)^2}  
\times \nonumber \\
&& \left( -2\,{{{m_{\gamma}}}^4} +
       2\,{{{m_{\gamma}}}^2}\,{{{q_0}}^2} - {{{q_0}}^4} -
       4\,{{{m_{\gamma}}}^2}\,{{\varepsilon }^2} -
       2\,{{{q_0}}^2}\,{{\varepsilon }^2}  \right. \nonumber \\
 &&    - 2\,{{\varepsilon }^4} +
       2\,\left( {{{m_{\gamma}}}^2} + {{\varepsilon }^2} \right) \,
        \left( {{{m_{\gamma}}}^2} - {{{q_0}}^2} +
          {{\varepsilon }^2} \right) \,\cos (2\,{q_0}\,T)  \nonumber \\
          &&\left. +
       4\,{q_0}\,\varepsilon \,
        \left( {{{m_{\gamma}}}^2} + {{\varepsilon }^2} \right) \,
        \sin (2\,{q_0}\,T) \right)
\end{eqnarray}
so that the renormalized delta function squared for this case becomes:

\begin{eqnarray}
&&-{1 \over 4 q_0^2 T^2 \left( {{{m_{\gamma}}}^2} + {{\varepsilon }^2} \right) ^2}
 \times \nonumber \\
&& \left( -2\,{{{m_{\gamma}}}^4} +
       2\,{{{m_{\gamma}}}^2}\,{{{q_0}}^2} - {{{q_0}}^4} -
       4\,{{{m_{\gamma}}}^2}\,{{\varepsilon }^2} -
       2\,{{{q_0}}^2}\,{{\varepsilon }^2}  \right. \nonumber \\
 &&    - 2\,{{\varepsilon }^4} +
       2\,\left( {{{m_{\gamma}}}^2} + {{\varepsilon }^2} \right) \,
        \left( {{{m_{\gamma}}}^2} - {{{q_0}}^2} +
          {{\varepsilon }^2} \right) \,\cos (2\,{q_0}\,T)  \nonumber \\
          &&\left. +
       4\,{q_0}\,\varepsilon \,
        \left( {{{m_{\gamma}}}^2} + {{\varepsilon }^2} \right) \,
        \sin (2\,{q_0}\, T) \right)
\end{eqnarray}
Here we have kept both a small photon mass (to regulate the infrared) as well as $
\varepsilon$.  For the brehmstrahlung contribution one can not set $\epsilon$ to zero.

At $m_{\gamma} \rightarrow 0$ this becomes:
\begin{equation}
{ \sin^2( q_0 T) \over  q_0^2 T^2}[1- {q_0^2 \over {\varepsilon}^2}] +
{q_0^2 \over 4 {\varepsilon}^4 T^2} + {1 \over \varepsilon^2 T^2}
\end{equation}

In order to obtain our previous results we used
$\varepsilon T -> \infty$. We now see to also have the unwanted terms going to zero we also
need
$\varepsilon^4 T ->
\infty$.
The actual rate of production of dileptons gets from this expression a contribution which
goes to zero as a constant divided by the total time. This constant is about the size of
the entire dilepton production rate at a $T= 20 fermis$. 
 In the pion gas case when $ T = 100 $  fermis, we need (in inverse units)
 $\varepsilon = 1$ for the brehmstrahlung contribution to be not that big a contamination, and for
$\epsilon$ to be small enough for the annhilation cross section to be reasonably accurate.
To take the limit $\varepsilon \rightarrow 0$ 
we choose for $ T > T_0 = 100$: 
\[  \epsilon  -> (T/T_0)^{-\delta} ; ~~~0 < \delta << 1, \]
in order to smoothly go to the covariant cutoff  result that is
\begin{equation}
{ \sin^2( q_0 T) \over  q_0^2 T^2}
\end{equation}

With this form for the $\varepsilon$, the finite $q_0$ dependent contribution
to the cross section goes to zero as $ 1 /T^{1-\delta}$.
Thus in  doing numerical simulations, we find that in order to avoid contamination
from brehmstrahlung processes we need to go to quite large hadronic time scales
$T > 1000$ to be in the asymptotic regime for the production of dileptons which
is an electromagnetic process.

For the brehmstrahlung process, the  effective $\delta$ function  is
independent of $k$ so that one can do the integration over $k$ for any $q_0$ to
obtain:

\begin{equation}
C_n = \Omega_n \int {dk \over (2 \pi)^n}  k^{n+1} {(1+n_k) n_k \over \omega_k^2}
\end{equation}
For $ \beta^{-1} =m_{\pi}$ we find
$C_1=  .127718$ ,  and  $C_3 = .0792387$ .

For the creation contribution we get a result similar to the annihilation
but with  $q_0 \rightarrow -q_0$, that is:

\begin{eqnarray}
F_{creation} &&=
 -{e^4  N_k^2 \over 4 q_0^2 \omega_k^2} \times \nonumber \\
&&
\{ {\sin^2 [(2 \omega_k+q_0) T] \over (q_0+2 \omega_k)^2 T^2}  +
{(-q_0+2 \omega_k)^2 \over 16 q_0^2 \omega_k^2 T^2 } \}
\end{eqnarray}
so that the renormalized square of the delta function is:
\[ \{ {\sin^2 [(2 \omega_k+q_0) T] \over (q_0+ 2 \omega_k)^2 T}  +
{(q_0-2 \omega_k)^2 \over 16 q_0^2 \omega_k^2 T} \}
\]
vs. the cutoff McLerran formula result:
\begin{equation}
{ \sin^2(\{ q_0+2 \omega_k \} T) \over  (q_0+ 2 \omega_k)^2 T^2}
\end{equation}

Using these formula, we have evaluated the dilepton production at fixed time
$T$ for both a pion gas and for the interacting field thory described
by the $\sigma$ model described above.  When the effective pion mass
goes negative, there is significant enhancement of the signal, however one
can see the finite time effects are still not controlled in our present
simulations.
In figure 8 we show the finite time effects for a pure pion gas where one
can determine the infinited time limit analytically. The time here is 100 fermis.
In figure 9 we see that quench conditions enhances significantly the production of low mass
dileptons over what one would find for a pure pion gas.
\epsfxsize=10cm
\epsfysize=10cm
\centerline{\epsfbox{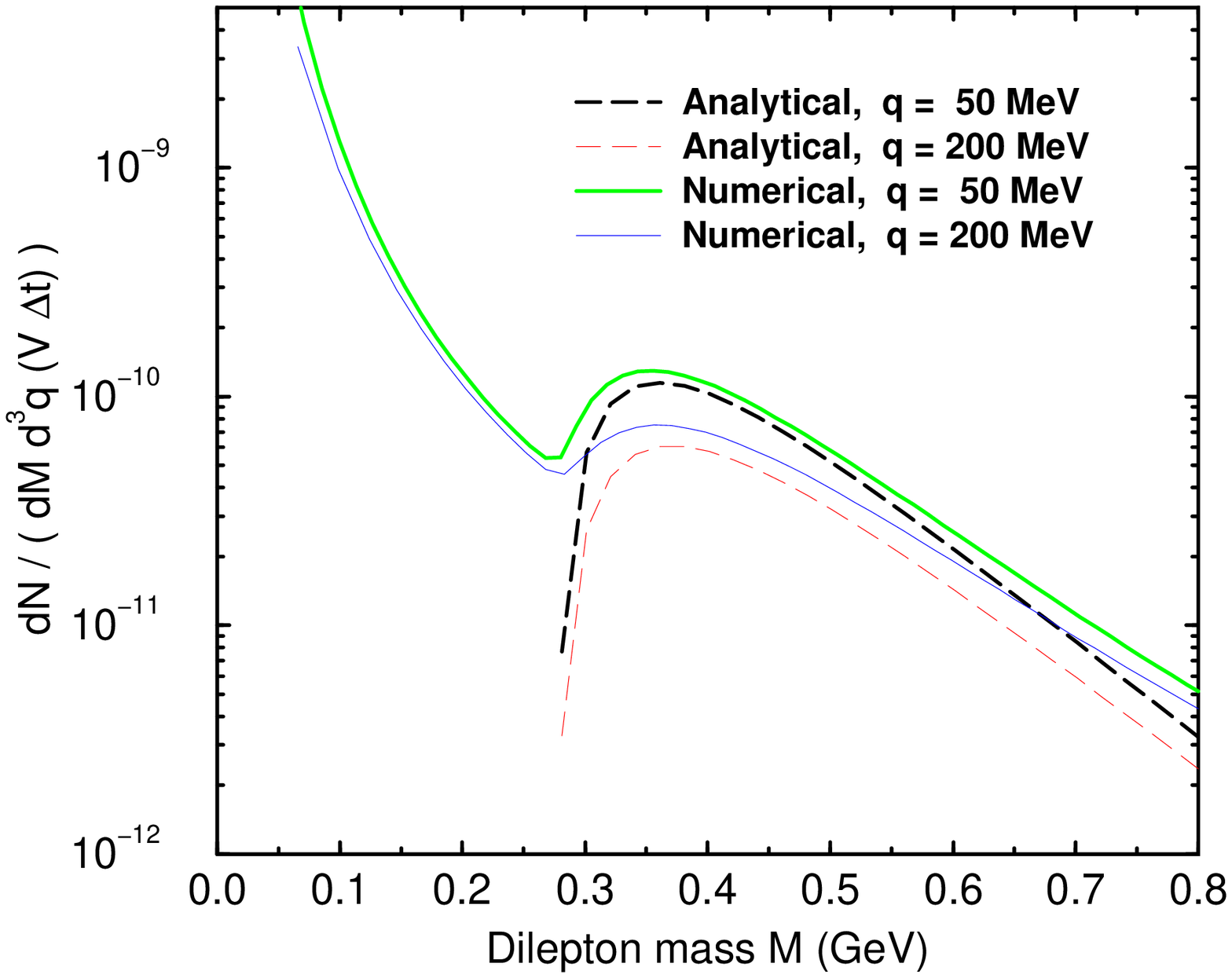}}
\vspace{.2cm}
FIG. 8. Finite Time effects for a pure pion gas. Here $t_f = 100f$
\\  
\epsfxsize=10cm
\epsfysize=10cm
\centerline{\epsfbox{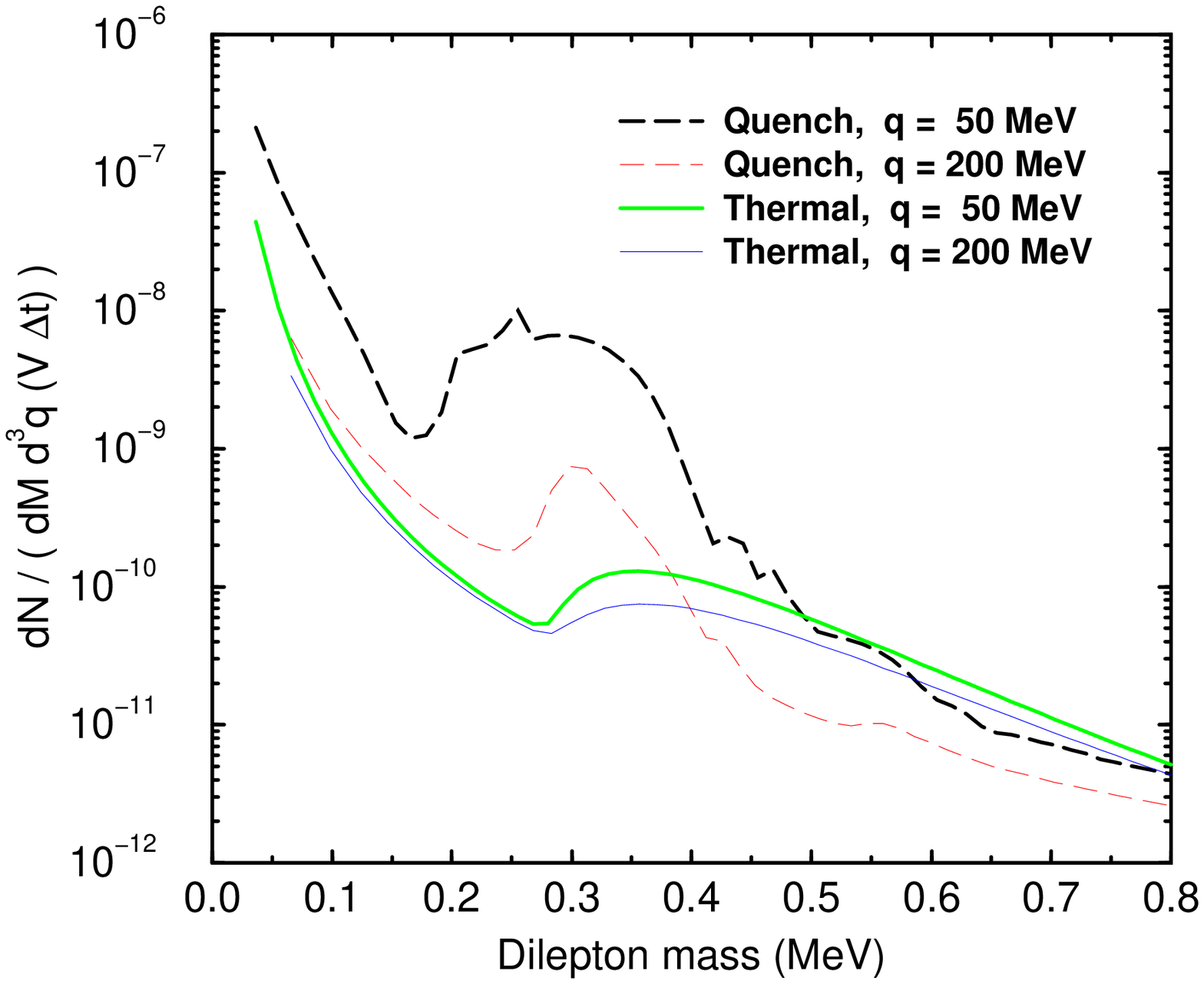}}
\vspace{.2cm}
FIG. 9. Comparison of quench (negative effective mass regime) vs. non-quench initial conditions for an interacting field theory described by the sigma model.
The large rise for small effective dilepton mass is the artifact of the 
finite time for the simulation. This effect slowly goes away as $1/t_f$.
\\

\section{Conclusions}
We have shown how to use the CTP formalism to calculate the dilepton 
spectrum arising from a time evolving or a thermal plasma.
For a plasma undergoing a chiral phase transition we expect a
strong signal for existence of DCC-states in $e^+ e^-$ - channel

\[ M_{inv} \sim 2 m_{\pi} \,\,\, ; q_{\perp} < 300 \, MeV \]

This would be visible by CERES if
  \[ k_{\perp} ^{cut} = 60\, MeV \]
In our calculations we have ignored possible important effects of 
direct two body scattering in the plasma which arise only in next order
in the $1/N$ approach. 
A similar enhancement seen by
D. Boyanovsky et. al.\cite{bib:boy} in photon spectrum.
Another problem for us is that our  result is influenced by large finite Time corrections  which are apparent for the free pion gas.

\section{ACKNOWLEDGEMENTS}  

The work presented here was done in 
collaboration with Emil Mottola, Yuval Kluger, Volker Koch, Juan Pablo Paz
Ben Svetitsky, Judah Eisenberg, Paul Anderson and John Dawson.
This work was supported by the Department of Energy.

\end{document}